\documentstyle[12pt]{article}
\include{psfig}
\newcommand{\quarkflow}[4]{\raisebox{-1.2ex}[3.0ex][3.0ex]{$%
   \begin{array}{r} 
	\mbox{\psfig{file=#1,height=6ex,width=2.4em}} \\*[-1ex]%
        {\scriptstyle #2\;#3\;#4\!} 
   \end{array}$}}

\def\slashmark#1#2#3{\global\setbox0=\hbox{\raise#2em
	\hbox{\kern#3em $#1\mathchar"0236$}}%
	\wd0=0pt \ht0=0pt \dp0=0pt \box0}
\def\vslash{{\mathchoice{\slashmark\displaystyle{-.1}{-.125}}%
		{\slashmark\textstyle{-.1}{-.125}}%
		{\slashmark\scriptstyle{-.075}{-.1}}%
		{\slashmark\scriptscriptstyle{-.06}{-.08}}v}}

\global\arraycolsep=2pt
\begin{document}

\begin{titlepage}

\begin{flushright}
UW/PT-96-07

\end{flushright}

\vspace{0.3cm}

\begin{center}
{\LARGE\bf Quenched Chiral Perturbation Theory for Baryons}
\end{center}

\vspace{0.8cm}

\begin{center}
{\large James N. Labrenz~~and~~Stephen R. Sharpe}\\
\medskip
{\em Department of Physics, University of Washington, Seattle, WA 98195}\\
\medskip
{(May 1996)}
\end{center}

\vspace{0.8cm}

\begin{small}
\centerline{ABSTRACT}
\medskip
We develop chiral perturbation theory for baryons in quenched QCD.
Quenching (the elimination of diagrams containing virtual quark loops)
is achieved by extending the Lagrangian method of Bernard and Golterman,
and is implemented in a theory where baryons are treated as fixed velocity
sources. Our method requires that the octet baryons be represented
by a three index tensor rather than by the usual matrix field.
We calculate the leading non-analytic corrections to the masses of octet 
and decuplet baryons. In QCD these are proportional to $M_\pi^3$.
We find that quenching alters the $M_\pi^3$ terms, but does not completely 
remove them. In addition, we find non-analytic contributions to baryon masses 
proportional to $M_\pi$ and $M_\pi^2 \log M_\pi$.
These terms, which are artifacts of quenching,
dominate over the $M_\pi^3$ terms for sufficiently small quark masses.
This pattern of corrections is different from that in most mesonic quantities,
where the leading non-analytic terms in QCD
(proportional to $M_\pi^4\log M_\pi$) are removed by quenching.
We also point out various pecularities of the quenched theory,
most notably that the $\Delta$ baryon can decay (if kinematically allowed),
in the sense that its two point function will be dominated at long Euclidean
times by a nucleon plus pion intermediate state.
\end{small}

\end{titlepage}

\section{Introduction and Summary}

A central goal of numerical simulations of lattice QCD is to
calculate the hadron spectrum and compare with experiment. 
Agreement between simulations and experiment would provide a crucial
test that QCD is the correct theory of the strong interactions
in the low energy domain.
Present simulations have yet to achieve this goal,\footnote{%
Opinions differ as to how close we are to reaching this goal.
For example, Ref. \cite{weingarten} is optimistic
while Ref. \cite{ourspect96} is more pessimistic.}
largely because of various approximations that must be made to
make the calculations computationally tractable.
Most important of these is the use of the quenched approximation,
in which the fermion determinant is approximated by a constant,
so that there are no internal quark loops.
The other significant approximations are the use of quarks with masses
heavier than their physical counterparts, and the use of
finite lattice spacing.

This paper is the first of two in which we study the importance
of the quenched approximation for baryon masses.
In this paper we develop chiral perturbation theory for baryons 
in the quenched approximation and calculate the dominant
contributions from one-loop graphs to both 
octet and decuplet baryon masses.
In the companion paper we work out some implications
of these results \cite{sharpebary}. 
In particular we use them to estimate the size of quenching errors in baryons, 
and show how they lead to a better understanding of how
to do chiral extrapolations in quenched QCD (QQCD).
This knowledge has already been useful for analyzing results of 
simulations \cite{ourspect96}.

To provide a context for the quenched results, it is useful to recall
the general form of the chiral expansion of baryon masses in QCD
(see for example Refs. \cite{jm,jenkins,bkm}),
\begin{equation}
M_{\rm bary} = M_0 + c_2 M_\pi^2 + c_3 M_\pi^3 + 
c_{4L} M_\pi^4 \log M_\pi + c_4 M_\pi^4 + \dots \,.
\end{equation}
Here $M_0$ is the mass in the chiral limit, and $M_\pi$ is a generic
pseudo-Goldstone boson mass. The dominant contributions come, in fact,
from the $K$ and $\eta$ mesons.
The $c_i$ are combinations of coefficients appearing in the chiral Lagrangian.
The expansion consists of terms which are analytic in the quark
masses---those proportional to $M_\pi^2 \propto (m_u+m_d)$ and
$M_\pi^4$---and non-analytic terms proportional to $M_\pi^3$ and
$M_\pi^4 \log M_\pi$.
The latter arise from infra-red divergences in loop graphs,
and thus are multiplied by constants ($c_3$ and $c_{4L}$) which are
determined in terms of leading order chiral coefficients.
For example, $c_3$ depends on the pion-nucleon couplings $F$ and $D$
(defined precisely below).
Thus, if one knows the lowest order chiral coefficients, one
can predict the form of the leading non-analytic terms.
The same is true for mesonic quantities (e.g. $M_\pi^2$ and $f_\pi$),
but with baryons the leading non-analytic terms contain one less power
of $M_\pi$ than the first analytic corrections
($M_\pi^3$ versus $M_\pi^4$), and are thus enhanced in the chiral limit,
while with mesons (as with the $c_{4L}$ term above) the enhancement
is only logarithmic.

In the quenched approximation we find that the expansion is different,
\begin{equation}
M_{\rm bary}^q = M_0^q + \delta\,c_1^q M_\pi 
+ \delta\,c_{2L}^q  M_\pi^2 \log M_\pi 
+ c_2^q M_\pi^2  + c_3^q M_\pi^3 +
c_{4L}^q M_\pi^4 \log M_\pi + c_4^q M_\pi^4 + \dots \,.
\label{eq:quexp}
\end{equation}
The constants are now different combinations of the coefficients
in the quenched chiral Lagrangian.
We have calculated the constants $c_1^q$, $c_{2L}^q$, $c_2^q$
and $c_3^q$, all of which are given in terms of lowest order
coefficients. We have not calculated $c_{4L}^q$, but for sufficiently
small $M_\pi$ (which, in numerical simulations, is adjustable)
the $M_\pi^4$ terms should be a small correction compared to the
lower order terms. 

The dimensionless constant $\delta$, which multiplies
the $c_1^q$ and $c_{2L}^q$ in Eq.~\ref{eq:quexp},
is a pure quenched artifact.
It appears also in mesonic quantities, 
and its likely magnitude is $\delta\approx 0.1$,
although there is no definitive determination \cite{rajanlat94}.
In Eq.~\ref{eq:quexp} it multiplies terms which will dominate the
corrections for small enough $M_\pi$.
If such terms are significant, then we should not trust the quenched results.
Fortunately, as discussed in the companion paper,
the numerical evidence suggests that these terms are small,
and so are largely curiosities.
They do, however, give another example of the sickness of the
quenched approximation in the chiral limit, a phenomenon first
pointed out for mesonic quantities in Refs. \cite{bg,sharpe1}.

Of more practical interest is the fact that the non-analytic terms
proportional to $M_\pi^3$ survive quenching, albeit multiplied by different
constants than in QCD. Naively, one might expect meson loops to require the
presence of underlying quark loops, and thus that quenching would remove all
the $M_\pi^3$ terms. This is indeed true for mesonic quantities,
where only artifacts proportional to $\delta$ survive.
Why this is not true for baryonic quantities is explained in Sec. \ref{sec:3}.

The presence of the $M_\pi^3$ terms has practical implications.
When doing chiral extrapolations to quenched results
one should fit to a form including these cubic terms.
It turns out that certain linear combinations
of baryon masses have expansions beginning at $O(M_\pi^3)$,
and these combinations can be used, in principle, 
to extract the constants $c_3^q$.
With these in hand, we can then make estimates of the errors
in baryon masses due to quenching.
The idea is to compare the size of the contributions of pion loops in
the two theories, and assume that any difference is a quenching error.
The uncertainty in these error estimates can be reduced by
forming suitable combinations of baryon masses.
Results from a pilot study along these lines
are presented in the companion paper.

In present simulations, the splitting between the octet and decuplet
baryons is substantially smaller than the pion mass,
$M_\Delta - M_N < M_\pi$. Thus in our calculations
we treat $M_\Delta-M_N$ as a small parameter,
and expand about the limit that the 
octet and decuplet baryons are degenerate. 
Eventually, simulations will be done in the opposite limit,
in which case the $\Delta$ should be treated as heavy
and integrated out of the effective theory.
As explained below, it is straightforward to adapt our results to this
new limit or, indeed, to any intermediate value of $(M_\Delta-M_N)/M_\pi$.
An interesting phenomenon which occurs when $M_\pi < M_\Delta - M_N$ is
that the quenched $\Delta$ decays. In Sec. \ref{sec:deltadecay},
we explain why this happens, and why it might not have been expected.

Finally, we note that the methods presented here can
be extended straightforwardly to ``partially quenched'' theories,
i.e. those in which there are internal quark loops, 
but the masses of the valence and loop quarks differ.

The outline of this paper is as follows. In the following section,
we develop the quenched chiral Lagrangian for baryons.
We then, in Sec. \ref{sec:3}, present the Feynman rules and sketch
a sample calculation. Section \ref{sec:4} contains our results
for baryon mass renormalizations.

\section{Chiral Lagrangian for Quenched Baryons}
\label{sec:2}

We calculate the quenched chiral corrections to the baryon masses 
using an effective Lagrangian of ``heavy'' baryons coupled to
the low-lying pseudoscalars.
The approach is a synthesis of two techniques which have
recently appeared in the literature:
1) Baryons are treated as fixed velocity sources,
thereby eliminating the baryon mass term from the Lagrangian
\cite{jm}; and
2) The theory is extended from QCD to QQCD by constructing 
a Lagrangian symmetric under the graded group
${\rm SU}(3|3)_{L} \times {\rm SU}(3|3)_{R}$ \cite{bg}.
The latter step corresponds to the addition
of ``bosonic quark'' degrees of freedom whose internal loops in
Feynman diagrams cancel those of the ordinary quarks.
\bigskip

\subsection{Review of results for QCD}

We start with the low-energy effective theory for QCD, in which
the pseudo-Goldstone bosons are coupled to octet and decuplet baryons.
The Goldstone fields are grouped in the usual $3\times3$ matrix $\pi(x)$
(normalized so that $\pi_{13} = K^{+}/\sqrt{2}$).
The octet baryons are similarly grouped in the $3\times3$ Dirac field $B(x)$ 
(with $B_{13}$ normalized to the proton).
Finally, the spin-$3/2$ baryons are represented by a 
Rarita-Schwinger tensor $T_{ijk}^\nu(x)$, symmetric in its three
flavor indices (with $T_{111}$ normalized to the $\Delta^{++}$).
Fixed velocity fields are defined by 
\begin{equation}
	B_v(x) = \exp(im_B\vslash v_\mu x^\mu) B(x)\,,\,\qquad
	T_v^\nu(x) = \exp(im_B\vslash v_\mu x^\mu) T^\nu(x)\,,
\label{eq:staticfield}
\end{equation}
where $v$ is the velocity, and $m_B$ is the mass of the
octet baryons in the chiral limit.
As shown in Ref. \cite{jm}, in the $m_B\to\infty$ limit,
the Dirac gamma matrix structure of the theory can be eliminated
in favor of $v_\mu$ and the spin-operators $S_v^\mu$.
We will explain the properties of $S_v^\mu$ as they are needed.

To construct a chirally invariant Lagrangian, we use fields
which have simple chiral transformation properties. These are the
exponentiated pion fields
\begin{equation}
 \Sigma(x) = \exp(2i\pi(x)/f)\,,
 \qquad 
 \xi(x) = \exp(i\pi(x)/f)\,,
\label{eq:defs1}
\end{equation}
the axial and vector currents
\begin{equation}
 A_\mu = i\mbox{\small$1\over2$}
	(\xi\partial_\mu\xi^\dagger - \xi^\dagger\partial_\mu\xi)\,,
 \qquad
 V_\mu = \mbox{\small$1\over2$}
	(\xi\partial_\mu\xi^\dagger + \xi^\dagger\partial_\mu\xi)\,,
\label{eq:defs2} 
\end{equation}
the mass terms
\begin{equation}
 {\cal M}^{\pm} = \xi^\dagger M \xi^\dagger \pm \xi M \xi\,,
 \qquad
M = {\rm diag}(m_u,m_d,m_s)\,,
\label{eq:defs3}
\end{equation}
and the covariant derivatives of baryon fields
\begin{eqnarray}
 {\cal D}^\mu B &=& \partial^\mu B + [V^\mu,B]\,, \\
 {\cal D}^\mu T_{ijk}^\nu &=& \partial^\mu T_{ijk}^\nu
 + (V^\mu)_i^{i'} T_{i'jk}^\nu + (V^\mu)_j^{j'} T_{ij'k}^\nu
 + (V^\mu)_k^{k'} T_{ijk'}^\nu \,.
\end{eqnarray}
Here, as in the following, we have dropped the subscript $v$ on the 
heavy baryon fields---it is always implicitly present.
Under ${\rm SU}(3)_{L} \times {\rm SU}(3)_{R}$ the meson fields transform as
\begin{eqnarray}
	\Sigma &\rightarrow& L\Sigma R^\dagger\,,
\label{eq:trans1} \\
	\xi &\rightarrow& L\xi U^\dagger(x) = U(x)\xi R^\dagger\,,
\label{eq:trans2} 
\end{eqnarray}
where $U(x)$ is defined implicitly through Eq.\ (\ref{eq:trans2}).
$A_\mu$, ${\cal M}^\pm$, $B_v$ and ${\cal D}^\mu B_v$, which are 
octets under the diagonal ${\rm SU}(3)$, all transform like
\begin{eqnarray}
	B_v(x) &\rightarrow& U(x) B_v U^\dagger(x) \,.
\label{eq:trans3}
\end{eqnarray}
Finally, the decuplet field and its covariant derivative both transform
as tensors, e.g.
\begin{eqnarray}
	T_{ijk}^\nu &\rightarrow& U_{i\ell} U_{jm} U_{kn} T_{\ell mn}^\nu \,.
\end{eqnarray}
Here, and in the following, we do not distinguish between
raised and lowered flavor indices. It will always be clear which
indices transform with $U$ and which with $U^\dagger$.

The most general chirally-symmetric Lagrangian can now be written down
as an expansion in momenta and quark masses.
The leading analytic and non-analytic corrections to the baryon
masses are obtained from the following terms
\begin{eqnarray}
  {\cal L} &=& {\cal L}_\pi + {\cal L}_{B\pi} + {\cal L}_{T\pi},
\nonumber\\ 
  {\cal L}_\pi &=& \mbox{\small$1\over4$} f^2
	\,{\rm tr}(\partial^\mu\Sigma\partial_\mu\Sigma^\dagger
		+ 2\mu\; {\cal M}^{+})\,, 
\label{eq:Lpi}\\ 
  {\cal L}_{B\pi} &=&  i\,{\rm tr}(\overline B v\!\cdot\!{\cal D} B)
\nonumber \\
	&+& 2D\; {\rm tr}(\overline B S^\mu\{A_\mu,B\})
		\;+\; 2F\; {\rm tr}(\overline B S^\mu[A_\mu,B]) \nonumber\\
	&+& 2\mu b_D\; {\rm tr}(\overline B \{{\cal M}^{+},B\})
		+ 2\mu b_F\; {\rm tr}(\overline B [{\cal M}^{+},B])
		+ 2\mu b_0\; {\rm tr}(\overline B B)\; 
				          {\rm tr}({\cal M}^{+}) \,,
\label{eq:LBpi}\\ 
  {\cal L}_{T\pi} &=& -i {\overline T}^\nu (v\!\cdot\!{\cal D}) T_\nu 
	         + \Delta M {\overline T}^\nu  T_\nu 
\nonumber \\
	&+& 2 {\cal H}\, {\overline T}^\nu S^\mu A_\mu T_\nu 
	+ {\cal C} \left( {\overline T}^\nu A_\nu B 
	+ \overline B A_\nu T^\nu\right)
\,. \nonumber \\
	&+& c {\overline T}^\nu {\cal M}^+ T_\nu - 
	    \overline\sigma {\overline T}^\nu T_\nu {\rm tr}({\cal M}^{+}) \,.
\label{eq:LTpi}
\end{eqnarray}
We have followed the notation of Jenkins and Manohar \cite{jm},
except for the octet mass terms
in which we follow Bernard {\em et al.} \cite{bkm}.
The traces involving the octet baryon are over flavor indices.
The contractions of flavor indices for terms involving decuplet baryons
are not shown explicitly---we discuss them for the quenched Lagrangian below.
Finally,
$\Delta M$ is the decuplet-octet mass splitting in the chiral limit.

\subsection{Quenched chiral Lagrangian for mesons}

We now consider the quenched theory. Bernard and Golterman have
developed a Lagrangian framework which provides a consistent
means for calculating the physics of the low-lying pseudoscalars
in QQCD \cite{bg}. 
We briefly review this technique, introduce a compatible representation
for the baryons, and then construct the quenched baryon Lagrangian 
analogous to that of Eqs.~(\ref{eq:LBpi},\ref{eq:LTpi}) above.

QQCD can be described by the addition of bosonic quark degrees of freedom, 
$\widetilde q_i$, one for each flavor of light quark. 
They have the same masses, one for one, as the original quarks, $q_i$. 
Integrating over the $\widetilde q_i$ in the functional integral
yields a determinant which exactly cancels that resulting from the
quark integration.
This theory is symmetric, at the classical level,
under the graded group ${\rm U}(3|3)_L \times {\rm U}(3|3)_R$,
and this symmetry dictates the form
of the low-energy effective theory for the pseudo-Goldstone bosons.
To construct this theory,
one replaces the field $\pi(x)$ in the definitions (\ref{eq:defs1})
by the field $\Phi(x)$, given in block matrix form as 
\begin{equation}
	\Phi = \left[ \begin{array}{cc}
		\pi & \chi^\dagger \\
		\chi & \widetilde\pi
		\end{array} \right] \,.
\end{equation}
Here $\pi$ contains the ordinary mesons $(q\overline q)$,
$\widetilde\pi$ the mesons composed of bosonic quarks
$(\widetilde q \overline{\widetilde q})$,
and $\chi$ and $\chi^\dagger$ the ``fermionic mesons''
($\widetilde q \overline q$ and $q \overline{\widetilde q}$  respectively).
The mass matrix is extended to
$M={\rm diag}(m_u,m_d,m_s,m_u,m_d,m_s)$, while the definitions of
axial and vector currents retain the same form (\ref{eq:defs2}).
The transformation properties of the fields are unchanged,
except that the matrices $L$, $R$ and $U$ are now elements of
$U(3|3)$.

Because of the anomaly, the chiral symmetry of the quantum theory is
reduced to the semidirect product 
$[{\rm SU}(3|3)_{L} \times {\rm SU}(3|3)_{R}]\otimes U(1)$.
The reduction in symmetry allows the quenched chiral Lagrangian 
to contain arbitrary functions of the field
\begin{equation}
	\Phi_0 = {\mbox{\footnotesize $1\over\sqrt3$}}{\rm str}\Phi
	\equiv{\mbox{\footnotesize $1\over\sqrt2$}}(\eta'-\widetilde\eta'),
\label{eq:etaprime}
\end{equation}
($\eta'$ is the usual ${\rm SU}(3)$ singlet meson, $\widetilde\eta'$ its ghostly
counterpart), since $\Phi_0$
is invariant under the quantum, though not the classical, symmetry group.
Putting this all together, one arrives at the following quenched
replacement for Eq.\ (\ref{eq:Lpi})
\begin{eqnarray}
   {\cal L}^{(Q)}_{\Phi} &=& {f^2\over4}
	\left[{\rm str}(\partial_\mu\Sigma\partial^\mu\Sigma^\dagger)\;V_1(\Phi_0)
	+2\mu\;{\rm str}({\cal M}^{+})\;V_2(\Phi_0)\right] 
\label{eq:Lqpi} \\
	&+&  \alpha_\Phi V_5(\Phi_0)\;\partial_\mu\Phi_0\partial^\mu\Phi_0
		-m_{0}^2\,V_0(\Phi_0)\;\Phi_{0}^2, \nonumber
\end{eqnarray}
where ${\rm str}$ denotes supertrace.
The potentials can be chosen to be even functions of $\Phi_0$, and are
normalized as $V_i(\Phi_{0}) = 1 + {\cal O}(\Phi_{0}^2)$. We will not, in fact,
need the higher order terms in these potentials.
Our notation follows that of Bernard and Golterman \cite{bg}, 
except that we use $m_0$ instead of $\mu$ (following Ref. \cite{sharpe1}) 
and $\alpha_\Phi$ instead of $\alpha$.
We also differ from \cite{bg} in choosing
the normalization of $f$ such that $f_\pi \approx 93\,$MeV.

At this point the development diverges from that in QCD.
In QCD, the last two terms in ${\cal L}^{(Q)}$ lead to wavefunction
renormalization and a mass shift for the $\eta'$. The $\eta'$ is then
heavy, and can be integrated out of the theory, giving the usual
Lagrangian ${\cal L}_\pi$ (\ref{eq:Lpi}).
In QQCD, by contrast, there is a cancelation between
diagrams with more than one insertion of either 
$\alpha_\Phi$ or $m_0$ on an $\eta'$ (or $\widetilde\eta'$) propagator
\cite{bg,sharpe1,bg-lat92}.
Thus the $\eta'$ and $\widetilde\eta'$ remain light,
and the $\alpha_\Phi$ and $m_0^2$ terms
must be included as interactions in ${\cal L}^{(Q)}$.
This leads to a more singular behavior of a number of quantities
(e.g. $m_\pi$ and $f_K$) in the chiral limit.
Furthermore, the new vertices destroy the usual power counting.
Higher loop diagrams involving $\alpha_\Phi$ and $m_0$ are 
{\em not} suppressed by powers of $p/\Lambda_\chi$ or $m_\pi/\Lambda_\chi$,
where $\Lambda_\chi\approx 1\,$GeV is the chiral cut-off.
We assume that $\alpha_\Phi$ and $m_0/\Lambda_\chi$ are small,
and work only to first order in these parameters.

Finally, although we have been talking about the $\eta'$, it is not, in fact,
a mass eigenstate unless the quarks are degenerate.
Instead, the flavor-neutral eigenstates are those with 
flavor composition $u\overline u$, $d\overline d$ and 
$s\overline s$, with squared masses $M_{qq}^2 = 2 \mu m_q$ ($q=u$, $d$, $s$).
These replace the $\pi_0$ and $\eta$ of QCD.
The flavor non-diagonal mesons are the same in both theories,
having the form $q_i \overline q_j$, with $M_{ij}^2=\mu (m_i + m_j)$.

\subsection{Baryon representations in quenched QCD}

To construct the chiral Lagrangian for baryons in QQCD,
we need to generalize the octet and decuplet fields $B$ and $T^\mu$. 
The corresponding quenched fields will contain 
additional baryons with compositions $qq\widetilde q$, 
$q\widetilde q\widetilde q$ and $\widetilde q\widetilde q\widetilde q$.
Even if we restrict the external states to be the usual $qqq$ baryons,
the extra baryons will appear in loops. 
Thus we must include these states in our effective Lagrangian. 
Just as when constructing the baryon part of the chiral Lagrangian in QCD,
all we need to know is how the states transform under the vector subgroup
${\rm SU}(3|3)_V$ (i.e. $L=R=U$).
The transformations under the full group are simply obtained by using
the position dependent $U(x)$ defined in Eq.\ (\ref{eq:trans2}).
Thus we need to determine the irreducible representations of ${\rm SU}(3|3)$
which, when restricted to the quark sector, 
contain only an octet or a decuplet of ${\rm SU}(3)$.

We begin with the octet, which is the more difficult case since it has
mixed symmetry. We construct representations using the ``quark'' field
$Q = (u, d, s, \widetilde u, \widetilde d, \widetilde s)$ 
and its conjugate $\overline{Q}$.
Under ${\rm SU}(3|3)_V$ these transform as fundamental and anti-fundamental
representations, respectively
\begin{equation}
	Q_i \longrightarrow U_{ij} Q_j \quad {\rm and} \quad
	\overline Q_i \longrightarrow \overline Q_j U_{ji}^\dagger\quad 
        {(i,j=1,6)}\,.
\label{eq:Qtrans}
\end{equation}
We now define the tensor spin-1/2 baryon field ${\cal B}_{ijk}(x)$
to have the same transformation properties as the following operator
constructed from the generalized quark fields
\begin{equation}
	{\cal B}_{ijk}^\gamma \sim
	\left[Q^{\alpha,a}_i Q^{\beta,b}_j Q^{\gamma,c}_k -
		Q^{\alpha,a}_i Q^{\gamma,c}_j Q^{\beta,b}_k \right]
		\varepsilon_{abc}(C\gamma_5)_{\alpha\beta}\,.
\label{eq:Bijk}
\end{equation}
Here $C=i\gamma_2\gamma_0$ is the charge conjugation matrix,
and $a$, $b$ and $c$ are color indices.
We have raised the color and spinor indices on the fields for the sake
of clarity, and suppressed the common position argument of all fields.
The transformation of ${\cal B}_{ijk}$ under $SU(3|3)_V$ is defined through
the r.h.s of Eq.\ (\ref{eq:Bijk})---the $Q$'s are first rotated ``inside''
the operator, and then the $U$'s are moved to the outside, giving
rise to a grading factor because the off-diagonal $3\times3$
blocks of $U$ are Grassman variables.
The result is
\begin{equation}
   {\cal B}_{ijk}^\gamma \longrightarrow
   (-)^{i'(j+j') + 
	(i'+j')(k+k')} U_{ii'}U_{jj'}U_{kk'} {\cal B}_{i'j'k'}^\gamma\,.
\label{eq:Btrans}
\end{equation}
Here we are using the following notation in the grading factor:
if the index on the field is ``anticommuting'' 
(in the range $1-3$) then the index equals 1 in the grading factor;
if the field index is ``commuting'' (in the range $4-6$) the
corresponding index equals 0 in the grading factor.

This construction of ${\cal B}$ automatically yields a representation
of $SU(3|3)_V$, since it is written in terms of $Q$'s.
The Dirac and color structure of the operator enforce the constraints
coming from the fact that ${\cal B}$ creates spin-1/2 baryons which are
color singlets.
The second term on the r.h.s. of Eq.\ (\ref{eq:Bijk}) is required
in order to make the representation irreducible.
To see this, consider the operator which result when
the indices $i-k$ are restricted to lie in the range $1-3$:
\begin{equation}
{\cal B}_{ijk}\Big|_R = {\cal B}^q_{ijk} + {\cal B}^q_{ikj}
\label{eq:reducingB}
\end{equation}
($R$ indicating restriction), where the usual quark baryon operator 
is defined to transform as in
\begin{equation}
{\cal B}^q_{ijk} \sim\left[q^{\alpha,a}_i q^{\beta,b}_j q^{\gamma,c}_k \right]
		\varepsilon_{abc}(C\gamma_5)_{\alpha\beta}\,.
\label{eq:threeqB}
\end{equation}
${\cal B}^q$ is anti-symmetric under $i\leftrightarrow j$, 
which allows the first two quark indices to be combined into an antiquark index
\begin{equation}
{\cal B}^q_{ijk} \propto \varepsilon_{ijk'} B^{k'}_k\,.
\label{eq:threetotwo}
\end{equation}
Now the problem is clear---the operator $B^{k'}_k$ creates both $SU(3)$
octets {\em and} singlets, but we only want the former.
The singlet is obtained by contracting ${\cal B}^q$ with $\varepsilon_{ijk}$.
It can be cancelled by symmetrizing the last two indices of ${\cal B}^q$,
as is done in Eq. (\ref{eq:reducingB}).
The octet part is, up to an overall constant, just the standard
$3\times3$ matrix used to represent baryons in the QCD chiral Lagrangian.
Using the normalizations discussed below, the explicit relation
between ${\cal B}^q$ and this field is
\begin{equation}
   {\cal B}_{ijk}\Big|_R = {\mbox{\footnotesize $1\over\sqrt6$}}
   \left( \varepsilon_{ijk'} B^{k'}_k + \varepsilon_{ikk'} B^{k'}_j\right)
	\,.
\label{eq:Bnorm}
\end{equation}
%

We have convinced ourselves that 
${\cal B}$ transforms in an irreducible representation of $SU(3|3)_V$.
It has dimension 70, and decomposes under $SU(3)$
as follows: an {\bf 8} each of $qqq$ and $\widetilde q\widetilde q\widetilde q$ states,
and a ${\bf 1} + {\bf 8} + {\bf 8} + {\bf 10}$ each of $qq\widetilde q$
and $q\widetilde q\widetilde q$ states.
It satisfies the symmetry properties (easily determined from the
above definitions)
\begin{eqnarray}
{\cal B}_{ijk} &=& (-)^{jk+1} {\cal B}_{ikj} \,, \nonumber\\
0 &=& {\cal B}_{ijk} + (-)^{ij+1} {\cal B}_{jik} 
	+ (-)^{ij+jk+ki+1} {\cal B}_{kji} \,.
\label{eq:Bsymm}
\end{eqnarray}
These relations show why ${\cal B}$ cannot be reduced to a 
two index form in QQCD---it has no grading independent
symmetry properties under the interchange of two indices. 

The need for a three index tensor makes sense also from another point
of view. The major aim of the formalism we are constructing is to allow
identification of contributions to mass renormalization which contain
internal quark loops. To do this we need to be able to follow
the flow of flavor through the Feynman diagrams. This is not straightforward
using the standard 2-index form for $B$, since one of the indices carries
the flavor of two quarks. Using the 3-index field ${\cal B}$, 
by contrast, there is a one-to-one correspondence 
between the terms which comprise a given diagram in QChPT 
and the flow of quark flavor.
In practice what happens is that all diagrams in which the quark
flow contains internal loops are cancelled by the corresponding diagrams
with $\widetilde q$ in the loops.

The representation containing the spin-3/2 baryons is simpler to
construct. The decuplet baryons are already represented by a symmetric
three-index tensor in QCD.
One only needs to extend the range of the indices, and apply the
appropriate symmetrization. We find that
\begin{equation}
{\cal T}^\mu_{\alpha,ijk} \sim
\left[  Q^{\alpha,a}_i Q^{\beta,b}_j  Q^{\gamma,c}_k
      + Q^{\beta,b}_i  Q^{\gamma,c}_j Q^{\alpha,a}_k
      + Q^{\gamma,c}_i Q^{\alpha,a}_j Q^{\beta,b}_k  \right]
\varepsilon_{abc} (C \gamma^\mu)_{\beta \gamma} \,.
\end{equation}
This transforms under ${\rm SU}(3|3)_V$ exactly as does ${\cal B}_{ijk}$, 
Eq.\ (\ref{eq:Btrans}).
It has the following symmetry properties (dropping Lorentz and Dirac
indices for clarity)
\begin{equation}
{\cal T}_{ijk} = (-)^{ij + 1} {\cal T}_{jik} = (-)^{jk+1} {\cal T}_{ikj} \,.
\label{eq:Tsymm}
\end{equation}
These imply that the representation is 38 dimensional,
containing a {\bf 10} of $qqq$ states, 
a {\bf 10} and an {\bf 8} of $qq\widetilde q$'s,
an {\bf 8} and a {\bf 1} of $q\widetilde q\widetilde q$'s,
and a lone {\bf 1} $\widetilde q\widetilde q\widetilde q$.
When the indices are restricted to the range $1-3$, the QQCD tensor is
proportional to the decuplet tensor used in QCD, and we choose normalizations
so that
\begin{equation}
{\cal T}_{ijk}\Big|_R = T_{ijk} \,.
\label{eq:Tnorm}
\end{equation}

\subsection{Quenched chiral Lagrangian for baryons}

To construct the quenched generalization of Eqs. (\ref{eq:LBpi})
and (\ref{eq:LTpi}) we need quantities which are invariant under 
${{\rm SU}(3|3)_{L} \times {\rm SU}(3|3)_{R}}$.
These we construct from the fields ${\cal B}$ and ${\cal T}$, 
their covariant derivatives, and their conjugates.
The covariant derivatives of both fields take the same form, exemplified by
\begin{equation}
 {\cal D}^\mu {\cal B}_{ijk} = \partial^\mu {\cal B}_{ijk}
 + (V^\mu)_{ii'} {\cal B}_{i'jk} + (-)^{i(j+j')}(V^\mu)_{jj'} {\cal B}_{ij'k}
 + (-)^{(i+j)(k+k')}(V^\mu)_{kk'} {\cal B}_{ijk'} \,,
\end{equation}
where $V^\mu$ is the vector-current defined by the
quenched generalization of Eq. (\ref{eq:defs2}).
The grading factors arise because,
in order to compensate for the position dependence of the $U$'s,
the $V$'s must act ``inside'' ${\cal B}$---i.e. as if they were
coupled directly to the generalized quark fields in Eq.~\ref{eq:Bijk}.
These covariant derivatives transform under
${{\rm SU}(3|3)_{L} \times {\rm SU}(3|3)_{R}}$
in the same way as the fields ${\cal B}$ and ${\cal T}$.
The conjugate baryon fields are defined by
\begin{eqnarray}
\overline{\cal B}_{kji}^{\gamma} &\sim&
	\left[\overline Q^{\gamma,c}_k 
	\overline Q^{\beta,b}_j \overline Q^{\alpha,a}_i -
	\overline Q^{\beta,b}_k \overline Q^{\gamma,c}_j 
		\overline Q^{\alpha,a}_i \right]
		\varepsilon_{abc}(C\gamma_5)_{\alpha\beta} \\
\overline{\cal T}^\mu_{kji,\alpha} &\sim&
\left[\overline Q^{\gamma,c}_k 
		\overline Q^{\beta,b}_j  \overline Q^{\alpha,a}_i 
      +\overline Q^{\alpha,a}_k 
		\overline Q^{\gamma,c}_j \overline Q^{\beta,b}_i
      +\overline Q^{\beta,b}_k  
		\overline Q^{\alpha,a}_j \overline Q^{\gamma,c}_i  \right]
\varepsilon_{abc} (C \gamma^\mu)_{\beta \gamma} \,,
\label{eq:BTbarijk}
\end{eqnarray}
and both transform in the same way, e.g.
\begin{equation}
   \overline{\cal B}_{kji}^\gamma \longrightarrow
   (-1)^{i'(j+j') + (i'+j')(k+k')}\ \overline{\cal B}_{i'j'k'}^\gamma
	U^{\dagger}_{k'k} U^{\dagger}_{j'j} U^{\dagger}_{i'i} \,.
\label{eq:Bbartrans}
\end{equation}

\smallskip
Using the group transformation properties, it is straightforward to
catalogue all the bilinear invariants. We use the compact notation
\smallskip
\begin{eqnarray}
  (\overline{\cal B}\Gamma{\cal B}) &\equiv& 
\overline{\cal B}_{kji}^{\alpha}\Gamma_{\alpha}\,\!^{\beta} 
		{\cal B}_{ijk,\beta}\,,
\label{eq:inv1}\\
  (\overline{\cal B}\Gamma A{\cal B}) &\equiv& 
\overline{\cal B}_{kji}^{\alpha}\Gamma_{\alpha}\,\!^{\beta}
				A_{ii'} {\cal B}_{i'jk,\beta}\,,
\label{eq:inv2}\\
  (\overline{\cal B}\Gamma{\cal B} A) 
	&\equiv& \overline{\cal B}_{kji}^{\alpha}\Gamma_{\alpha}\,\!^{\beta}
                                A_{kk'} {\cal B}_{ijk',\beta}
	\times (-)^{(i+j)(k+k')}\,,
\label{eq:inv3}\\
%
  (\overline{\cal T}^\mu\Gamma{\cal T}_\mu) &\equiv&
  \overline{\cal T}^\mu_{kji,\alpha} \Gamma^\alpha\,\!_{\beta} 
	{\cal T}_{\mu,ijk}^\beta \,,
\label{eq:Tinv1}\\
  (\overline{\cal T}^\mu\Gamma A^\nu{\cal T}_\mu) &\equiv&
 \overline{\cal T}^\mu_{kji,\alpha} \Gamma^\alpha\,\!_{\beta} 
                            A_{ii'}^\nu{\cal T}_{\mu,i'jk}^\beta \,,
\label{eq:Tinv2}\\
  (\overline{\cal B} \Gamma A^\mu T_\mu) &\equiv&
 \overline{\cal B}_{kji}^{\alpha} \Gamma^\alpha\,\!_{\beta} 
                            A_{ii'}^\mu{\cal T}_{\mu,i'jk}^\beta \,.
\label{eq:Tinv3}
\end{eqnarray}
There is also the conjugate of the last quantity, i.e.
$(\overline{\cal T}^\mu \Gamma A_\mu {\cal B})$.
In the above constructions,
$A$ is an operator transforming like the axial current $A_\mu$, 
and $\Gamma$ is an arbitrary Dirac matrix. 
In fact, the change to fixed-velocity fields simplifies the Dirac structure
as for QCD, and in practice only the
spin operator $S_\mu$ enters in place of $\Gamma$. 
Various possible terms are absent from the list above because they
can be rewritten in the above form using the symmetries 
of ${\cal B}$ and ${\cal T}$. 
For example, a $(\overline {\cal B} A {\cal B})$ bilinear with $A$
coupling to the second index of ${\cal B}_{ijk}$ is redundant
due to the symmetry ${\cal B}_{ijk} = (-1)^{jk+1} {\cal B}_{ikj}$. 
Other terms simply vanish, e.g. constructions involving 
${\cal B}_{kji} T_{ijk}$.

We can now write down the relevant part of the quenched Lagrangian
for baryons. 
It consists of 
\begin{eqnarray}
  {\cal L}^{(Q)}_{{\cal B}\Phi}& = & 
	i(\overline {\cal B} v\!\cdot\!{\cal D} {\cal B})
\nonumber \\ 
	&+& 2\alpha\; (\overline{\cal B} S^\mu{\cal B} A_\mu)
	\;+\; 2\beta\; (\overline{\cal B} S^\mu A_\mu{\cal B})
	\;+\; 2\gamma\; (\overline{\cal B} S^\mu{\cal B}) {\rm str}(A_\mu)
\label{eq:LqBpi} \\
	&+& \alpha_M\; (\overline{\cal B}{\cal B}{\cal M}^{+})
	\;+\; \beta_M\; (\overline{\cal B}{\cal M}^{+}{\cal B})
	\;+\; \sigma\; (\overline{\cal B}{\cal B}){\rm str}({\cal M}^{+})
\nonumber
\end{eqnarray}
and
\begin{eqnarray}
 {\cal L}^{(Q)}_{{\cal T}\Phi} &=&
	\;-\; i ({\overline {\cal T}}^\nu (v\!\cdot\!{\cal D}) {\cal T}_\nu)
        \;+\; \Delta M ({\overline {\cal T}}^\nu  {\cal T}_\nu) 
\nonumber \\
	&+& 2\,{\cal H}\, ({\overline {\cal T}}^\nu S^\mu A_\mu {\cal T}_\nu) 
	\;-\; \sqrt{\mbox{\small$3\over2$}}\, {\cal C} 
		\left[ ({\overline {\cal T}}^\nu A_\nu {\cal B}) 
		+ (\overline {\cal B} A_\nu {\cal T}^\nu) \right]
	\;+\; 2\,\gamma' (\overline {\cal T}^\nu S^\mu 
			{\cal T}_\nu) {\rm str}(A_\nu)
\label{eq:LqTpi}\\
	&+& c\, (\overline {\cal T}^\nu {\cal M}^+ {\cal T}_\nu) 
	\;-\; \overline\sigma\, (\overline {\cal T}^\nu {\cal T}_\nu)\,
		 {\rm str}({\cal M}^{+}) \,.
\nonumber
\end{eqnarray}
Each term can be multiplied by an arbitrary even function of
$\Phi_0$, but the higher order vertices these functions produce 
do not contribute at the order we are working.
There are also terms proportional to
$(\overline{\cal B}{\cal B})\Phi_{0}^2$
and $(\overline{\cal B}{\cal B})\partial_\mu\Phi_{0}\partial^\mu\Phi_{0}$,
but these do not contribute to mass renormalization at one-loop. 
In quark-flow language, 
these are double hairpin vertices and lead to closed quark loops \cite{bg}.
Finally we note that ${\rm str}(A_\nu)$ 
is proportional to $\partial_\nu \Phi_0$,
so possible additional terms involving the latter are not independent.

The quenched Lagrangian (\ref{eq:LqBpi})-(\ref{eq:LqTpi}) 
looks very similar in form to that for QCD, 
Eqs. (\ref{eq:LBpi})-(\ref{eq:LTpi}).
To make the correspondence precise we need to pick out the parts of
${\cal L}^{(Q)}_{{\cal B}\Phi}$ and ${\cal L}^{(Q)}_{{\cal T}\Phi}$ 
containing only $qqq$ fields. 
This is straightforward for the terms involving the spin-3/2 field,
since both quenched and full fields have three indices, so the
structure of the allowed contractions is the same.
For example, using Eq.\ (\ref{eq:Tnorm}), one finds that
\begin{equation}
({\overline {\cal T}}^\nu S^\mu A_\mu {\cal T}_\nu)\Big|_R
 = {\overline T}^\nu S^\mu A_\mu T_\nu\,.
\end{equation}
Thus the coefficients $\Delta M$, $c$ and ${\cal H}$ play the same role
in the quenched Lagrangian as they do in the full theory.
It is important to realize, however, that there is no reason for the
coefficients to have the same values in the two theories.
Despite this, we use the same symbols so that the
physical significance of each term can be more easily recognized.

Terms involving the spin-1/2 field require more work to interpret.
We need to convert from the three index tensor ${\cal B}$ to the matrix $B$.
Using Eq.\ (\ref{eq:Bnorm}) we find
\begin{eqnarray}
	(\overline{\cal B}{\cal B})\Big|_R &=& {\rm tr}(\overline B B)\,,
\label{eq:corresp1}\\
	(\overline{\cal B}{\cal B} A)\Big|_R &=& 
	\mbox{\small$2\over3$}{\rm tr}(\overline BAB)
	+\mbox{\small$1\over6$}{\rm tr}(\overline BB){\rm tr}(A)
	-\mbox{\small$1\over6$}{\rm tr}(\overline BBA)\,,
\label{eq:corresp2}\\
	(\overline{\cal B} A{\cal B})\Big|_R &=& 
	-\mbox{\small$1\over3$}{\rm tr}(\overline BAB)
        +\mbox{\small$2\over3$}{\rm tr}(\overline BB){\rm tr}(A)
	-\mbox{\small$2\over3$}{\rm tr}(\overline BBA) \,,
\label{eq:corresp3}\\
	(\overline{\cal B}{\cal B}){\rm str}(A)\Big|_R 
	&=& {\rm tr}(\overline B B){\rm tr}(A) \,.
\label{eq:corresp4}\\
({\overline {\cal T}}^\nu A_\nu {\cal B})\Big|_R
 &=& - \sqrt{\mbox{\small$2\over3$}} {\overline T}^\nu A_\nu B 
 = - \sqrt{\mbox{\small$2\over3$}} 
	{\overline T}^\nu_{ijk} A_{\nu,ii'} B_{jj'} \varepsilon_{i'j'k}\,. 
\label{eq:corresp5}
\end{eqnarray}
The first of these relations 
is actually the condition which sets the normalization in (\ref{eq:Bnorm}). 
From Eq.~(\ref{eq:corresp5}) we see that the coefficient ${\cal C}$ has the
same significance in quenched and full QCD.
As for the other coefficients, the quenched Lagrangian is equal to
that for QCD, Eq.\ (\ref{eq:LBpi}), if we make the identifications
\begin{eqnarray}
\begin{array}{lll}
	\phantom{_M}\alpha = 2({1\over3}D +F)\,,\quad &
	\phantom{_M}\beta = (-{5\over3}D +F)\,,\quad &
	\gamma = 2(D-F)\,,
\\*[+1ex]
	\alpha_M = 4({1\over3}b_D + b_F)\mu\,,\quad &
	\beta_M = 2(-{5\over3}b_D + b_F)\mu\,,\quad &
	\sigma = 2(b_0 +b_D -b_F)\mu\,.\phantom{aa} \end{array}
\label{eq:Qparameters}
\end{eqnarray}
With the exception of the result for $\gamma$, we use these
relations to reexpress our quenched results in terms of $D$, $F$, $b_D$, $b_F$
and $b_0$, rather than $\alpha$, $\beta$, $\alpha_M$, $\beta_M$ and $\sigma$.
This allows a more direct comparison with results from
chiral perturbation theory in QCD.
We reiterate, however, that the values of the coefficients will be different
in the two theories.

With all these correspondences in hand, we can now see the most
important difference between the Lagrangians in QCD and QQCD, namely that
the latter has two additional coefficients, $\gamma$ and $\gamma'$.
Eq.\ (\ref{eq:Qparameters}) shows that $\gamma$ is non-zero in QCD,
but that it is not independent of $D$ and $F$. 
In QQCD, by contrast, $\gamma$ is an independent parameter.
Both the new terms involve a baryon bilinear coupled to ${\rm str}(A_\nu)
\propto \partial_\nu \Phi_0$, and thus represent couplings of
the $\eta'$ and $\widetilde\eta'$ to baryons. 
These are independent couplings because there is no symmetry 
connecting the couplings of ${\rm SU}(3)$ octet and singlet mesons.
In the quark-flow language, these are ``hairpin'' vertices---%
the $q$ and $\overline q$ (or $\widetilde q$ and $\overline{\widetilde q}$) 
in the $\eta'$ (or $\widetilde\eta'$) annihilate.
Such couplings are present in QCD, but are absent from the chiral Lagrangian
because the $\eta'$ is heavy and can be integrated out.

\section{Flavor Feynman rules and a sample calculation}
\label{sec:3}

Calculations of Feynman graphs separate into  ``Dirac'' and flavor parts. 
The former are standard, using the Feynman rules and methodology
worked out in Refs. \cite{jm,jenkins}.
We do not give further details here.
The flavor part of the calculation is new,
and we have developed a diagrammatic method for carrying it out.
This method not only simplifies the calculations, 
but also allows one to trace the underlying quark flows.
In this section we explain our method, and sketch a sample calculation.

\subsection{Feynman rules}

The quenched baryon Lagrangian is written in terms of the 
three index tensor fields ${\cal B}_{ijk}$ and ${\cal T}_{ijk}$, 
each of which has 216 components.
The symmetry relations [Eqs.~(\ref{eq:Bsymm}) and (\ref{eq:Tsymm})] 
reduce the number of independent fields to 70 and 38, respectively. 
One could proceed by expressing the Lagrangian in terms of these 
independent fields.
By construction the baryon propagators would then be
diagonal and normalized.
The vertices, on the other hand, would involve complicated flavor factors.
In practice, we find it more useful to reexpress the Feynman rules 
in terms of all of the components of ${\cal B}$ and ${\cal T}$,
and impose the relations between the components only on external lines.
As we show explicitly below,
the symmetry relations are automatically maintained on internal lines 
by the structure of the baryon propagators.

We are interested in diagrams in which the external baryons
are $qqq$ states. Thus, we need to display the independent fields
explicitly when the indices of ${\cal B}$ and ${\cal T}$ lie in
the range $1-3$. Using eqs.~(\ref{eq:Bnorm}), the octet baryons are
\begin{equation}
\begin{array}{lll}
	p=b_{12}\,,\;\;          & n=b_{21}\,,\;\; & \\ 
	\Sigma^+=b_{13}\,,\;\;   & \Sigma^0=b_{(+)}\,,\;\;  
	& \Sigma^-=b_{23}\,, \\
	\Xi^0=b_{31}\,,\;\;      & \Xi^-=b_{32}\,,\;\; \\
	\Lambda^0=b_{(-)}\,.\;\; & & \end{array}
\label{eq:states}
\end{equation}
where we have used the symmetry relations to define
\begin{eqnarray}
b_{ij} &\equiv& \sqrt{6}\, {\cal B}_{iij}
		= \sqrt{6}\, {\cal B}_{iji}
		= -\sqrt{ \mbox{\small$3\over2$}}\,
			{\cal B}_{jii} \qquad (i\not=j),
\label{eq:bij}\\*[+1ex]
b_{(+)} &\equiv& \sqrt3\, ({\cal B}_{123}+{\cal B}_{231})
	= -\sqrt3\, {\cal B}_{312}
	= \sqrt3\, ({\cal B}_{132}+{\cal B}_{213})
	= -\sqrt3\, {\cal B}_{321}, \qquad
\label{eq:bplus}\\*[+1ex]
b_{(-)} &\equiv& ({\cal B}_{123} - {\cal B}_{231})
	= ({\cal B}_{132} - {\cal B}_{213})\,.
\label{eq:bminus}
\end{eqnarray}
For the decuplet baryons the relation to physical fields is just
as for the usual tensor field, for example 
\begin{equation}
\begin{array}{ll}
\Delta^{++} = {\cal T}_{111}\,,\quad&
\Sigma^{*,\,+} = \sqrt3\, {\cal T}_{113}
	     = \sqrt3\, {\cal T}_{131}
	     = \sqrt3\, {\cal T}_{311} \,, \\
\Omega^{-}  = {\cal T}_{333} \,, &
\Xi^{*,\,0} = \sqrt3\, {\cal T}_{133}
	    = \sqrt3\, {\cal T}_{313}
	    = \sqrt3\, {\cal T}_{331} \,. \end{array}
\label{eq:tstates}
\end{equation}

We now present the Feynman rules, beginning with the ``pion'' propagator
\begin{equation}
%
%
	\Phi_{m}^{n} \!\!\!
	\raisebox{-0.9ex}{\psfig{file=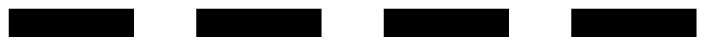,height=5ex,width=5em}} \;
	\Phi_{i}^{j} \phantom{aa} = \phantom{aa}
	{i\over p^2 - M_{ij}^2}\; \times\;  {1\over2}\, (-)^{in+1}
	\begin{array}{c}  
	\begin{array}{lr} 
	{\scriptstyle n}&{\scriptstyle m}\end{array}\\*[-.2ex]
	\mbox{\psfig{file=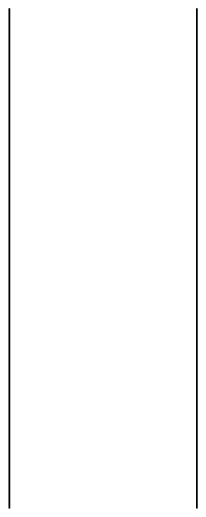,height=5ex,width=.6em}} \\*[-1ex]
	\begin{array}{lr} {\scriptstyle i}&{\scriptstyle j}
	\end{array} 
	\end{array} \,.\\
\end{equation}
Here, lines joining indices denote Kronecker deltas, 
so that the flavor factor is $\delta_{ni}\delta_{mj}$. 
The form of the grading factor follows from the
the properties of the supertrace.

To derive the baryon propagators, 
we write each component of ${\cal B}$ or ${\cal T}$ 
in terms of the independent fields, 
and then use the fact that the propagator is diagonal for these fields.
A useful trick for obtaining the relative factors, and particularly
the grading factors, is to think in terms of the underlying $Q$ fields.
For example, for the spin-1/2 propagator, we rewrite the field as
\begin{equation}
	{\cal B}_{ijk} \sim 
	[ Q_iQ_jQ_k - (-1)^{jk}Q_iQ_kQ_j ] 
		\times {\rm spin}\times {\rm color},
\end{equation}
insert this relation and its conjugate into the 
propagator $\langle{\cal B}_{lmn} \overline{\cal B}^{ijk}\rangle$, 
and then compute the ordinary Wick contractions of the $Q$ vectors. 
Reordering the $Q$'s to form 
singlet combinations results in a common spin and color factor
multiplied by a constant which includes the grading factors.
In this way we obtain
\begin{eqnarray}
%
	{\cal B}_{lmn}\!\!\!
	\raisebox{-0.8ex}{\psfig{file=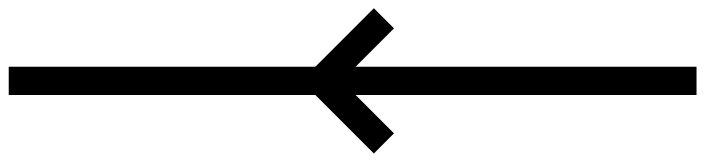,height=5ex,width=5em}} \;
	\overline{\cal B}_{kji} \phantom{aa} &\phantom{|} = & 
	{i\over (v\!\cdot\! k)+i\epsilon}\, {\cal F}^{(1/2)}_{lmn,ijk} \,,
\label{eq:BBbar} \\
{\cal F}^{(1/2)}_{lmn,ijk}
&\phantom{|} = &
 {1\over6} \left[ \;
2\begin{array}{c}
 \begin{array}{lcr}
	{\scriptstyle l}&{\scriptstyle m}&{\scriptstyle n}
 \end{array} \\*[-.2ex]
	\mbox{\psfig{file=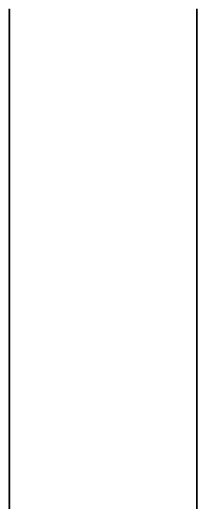,height=5ex,width=1.2em}} \\*[-1ex]
 \begin{array}{lcr} 
	{\scriptstyle i}&{\scriptstyle j}&{\scriptstyle k}
 \end{array} 
 \end{array} 
-\; 2 (-)^{jk}
 \begin{array}{c}
 \begin{array}{lcr} 
	{\scriptstyle l}&{\scriptstyle m}&{\scriptstyle n}
 \end{array} \\*[-.2ex]
	\mbox{\psfig{file=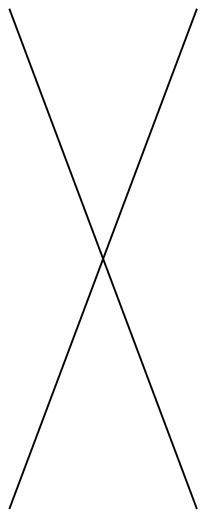,height=5ex,width=1.2em}} \\*[-1ex]
 \begin{array}{lcr} 
	{\scriptstyle i}&{\scriptstyle j}&{\scriptstyle k}
 \end{array} 
 \end{array} 
+\; (-)^{ij}
 \begin{array}{c}
 \begin{array}{lcr} 
	{\scriptstyle l}&{\scriptstyle m}&{\scriptstyle n}
 \end{array} \\*[-.2ex]
	\mbox{\psfig{file=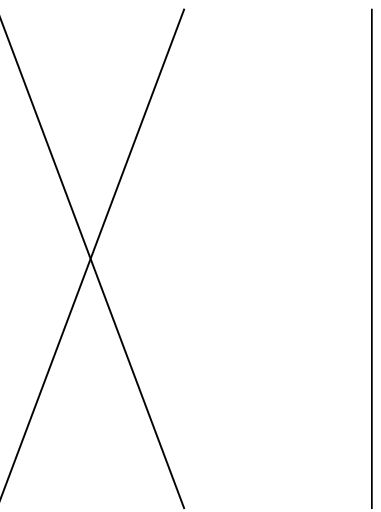,height=5ex,width=1.2em}} \\*[-1ex]
 \begin{array}{lcr} 
	{\scriptstyle i}&{\scriptstyle j}&{\scriptstyle k}
 \end{array} 
 \end{array} 
\right. \label{eq:Bflavorprop} \\
&\phantom{|}& \mbox{}\phantom{xx} 
-\; (-)^{ik+jk} \left.
 \begin{array}{c}
 \begin{array}{lcr} 
	{\scriptstyle l}&{\scriptstyle m}&{\scriptstyle n}
 \end{array} \\*[-.2ex]
	\mbox{\psfig{file=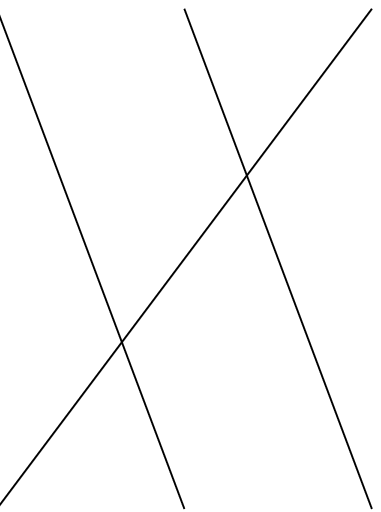,height=5ex,width=1.2em}} \\*[-1ex]
 \begin{array}{lcr} 
	{\scriptstyle i}&{\scriptstyle j}&{\scriptstyle k}
 \end{array} 
 \end{array} 
-\; (-)^{ij+ik} 
 \begin{array}{c}
 \begin{array}{lcr} 
	{\scriptstyle l}&{\scriptstyle m}&{\scriptstyle n}
 \end{array}\\*[-.2ex]
	\mbox{\psfig{file=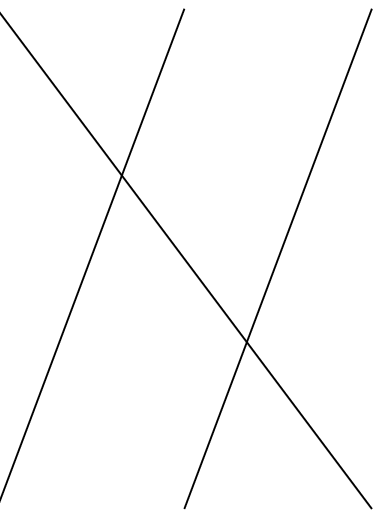,height=5ex,width=1.2em}} \\*[-1ex]
 \begin{array}{lcr} 
	{\scriptstyle i}&{\scriptstyle j}&{\scriptstyle k}
 \end{array} 
 \end{array} 
+\; (-)^{ij+ik+jk}
 \begin{array}{c}
 \begin{array}{lcr} 
	{\scriptstyle l}&{\scriptstyle m}&{\scriptstyle n}
 \end{array}\\*[-.2ex]
	\mbox{\psfig{file=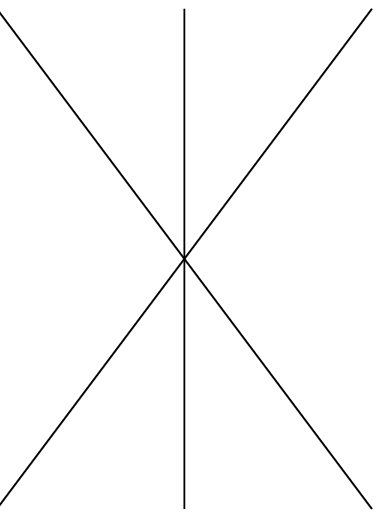,height=5ex,width=1.2em}} \\*[-1ex]
 \begin{array}{lcr} 
	{\scriptstyle i}&{\scriptstyle j}&{\scriptstyle k}
 \end{array} 
 \end{array}
\right]\,, \nonumber \\
%
%
	{\cal T}^\nu_{lmn}\!\!\!
	\raisebox{-0.8ex}{\psfig{file=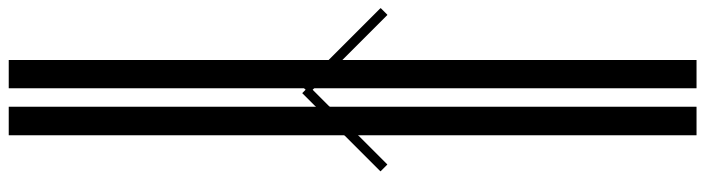,height=5ex,width=5em}} \;
	\overline{\cal T}^\mu_{kji} \phantom{aa} &= \phantom{|}&
	{i P^{\nu\mu}\over (v\!\cdot\! k)+i\epsilon}\ 
		{\cal F}^{(3/2)}_{lmn,ijk}
 \label{eq:TTBar} \\
{\cal F}^{(3/2)}_{lmn,ijk} &\ =\ \phantom{|}& 
{1\over6} \left[ \;
 \begin{array}{c}
 \begin{array}{lcr}
	{\scriptstyle l}&{\scriptstyle m}&{\scriptstyle n}
 \end{array} \\*[-.2ex]
	\mbox{\psfig{file=BBbar1.ps,height=5ex,width=1.2em}} \\*[-1ex]
 \begin{array}{lcr} 
	{\scriptstyle i}&{\scriptstyle j}&{\scriptstyle k}
 \end{array}
 \end{array} 
-\; (-)^{jk}
 \begin{array}{c}
 \begin{array}{lcr} 
	{\scriptstyle l}&{\scriptstyle m}&{\scriptstyle n}
 \end{array} \\*[-.2ex]
	\mbox{\psfig{file=BBbar2.ps,height=5ex,width=1.2em}} \\*[-1ex]
 \begin{array}{lcr} 
	{\scriptstyle i}&{\scriptstyle j}&{\scriptstyle k}
 \end{array}
 \end{array} 
-\; (-)^{ij}
 \begin{array}{c}
 \begin{array}{lcr} 
	{\scriptstyle l}&{\scriptstyle m}&{\scriptstyle n}
 \end{array} \\*[-.2ex]
	\mbox{\psfig{file=BBbar3.ps,height=5ex,width=1.2em}} \\*[-1ex]
 \begin{array}{lcr} 
	{\scriptstyle i}&{\scriptstyle j}&{\scriptstyle k}
 \end{array} 
 \end{array} 
\right. \label{eq:Tflavorprop} \\
&\phantom{|}& \phantom{xx} 
+\; (-)^{ij+ik} \left.
 \begin{array}{c}
 \begin{array}{lcr} 
	{\scriptstyle l}&{\scriptstyle m}&{\scriptstyle n}
 \end{array} \\*[-.2ex]
	\mbox{\psfig{file=BBbar4.ps,height=5ex,width=1.2em}} \\*[-1ex]
 \begin{array}{lcr} 
	{\scriptstyle i}&{\scriptstyle j}&{\scriptstyle k}
 \end{array} 
 \end{array} 
+\; (-)^{ik+jk} 
 \begin{array}{c}
 \begin{array}{lcr} 
	{\scriptstyle l}&{\scriptstyle m}&{\scriptstyle n}
 \end{array} \\*[-.2ex]
	\mbox{\psfig{file=BBbar5.ps,height=5ex,width=1.2em}} \\*[-1ex]
 \begin{array}{lcr} 
	{\scriptstyle i}&{\scriptstyle j}&{\scriptstyle k}
 \end{array}
 \end{array} 
-\; (-)^{ij+ik+jk}
 \begin{array}{c}
 \begin{array}{lcr} 
	{\scriptstyle l}&{\scriptstyle m}&{\scriptstyle n}
 \end{array} \\*[-.2ex]
	\mbox{\psfig{file=BBbar6.ps,height=5ex,width=1.2em}} \\*[-1ex]
 \begin{array}{lcr} 
	{\scriptstyle i}&{\scriptstyle j}&{\scriptstyle k}
 \end{array}
 \end{array}
\right]\,. \nonumber
\end{eqnarray}
Here $P^{\mu\nu} = v^\mu v^\nu - g^{\mu\nu} - (4/3) S^\mu S^\nu$ is the
spin-3/2 projection matrix, and the ${\cal F}^{(j)}$ are flavor
projectors, with $j$ labeling the spin. We discuss the properties of
the ${\cal F}$'s below.

We obtain the vertices in a similar fashion. The baryon-pion vertices are
\begin{eqnarray}
&\phantom{|}&
  \overline{\cal B}_{nml} \!\!
  \begin{array}{c} 
	\Phi_{p}\,^{q} \\
	\mbox{\psfig{file=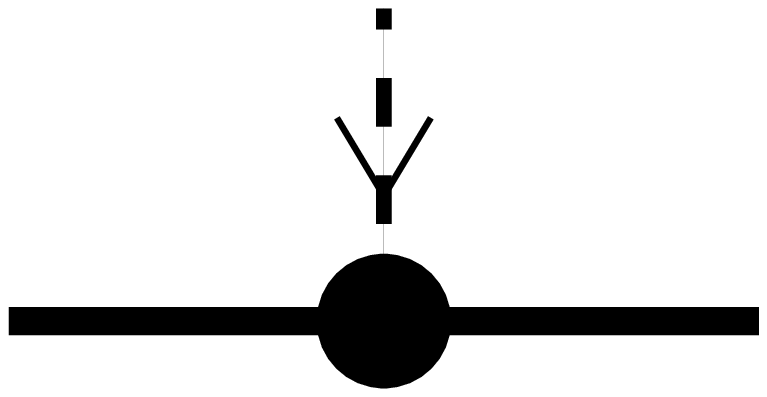,height=5em,width=5em}}
  \end{array}
  {\cal B}_{ijk}  \phantom{aa} = \nonumber \\
&\phantom{|}& \phantom{xxxxxxxx} 
{k \cdot S \over f} \left[ \;
2 \alpha (-)^{(i+j)(p+q)}\,
 \begin{array}{l}
 \begin{array}{llllr}
	{\scriptstyle l}&{\scriptstyle m}&{\scriptstyle n}&
	{\scriptstyle p}&{\scriptstyle q}
 \end{array} \\*[-.2ex]
	\mbox{\ \psfig{file=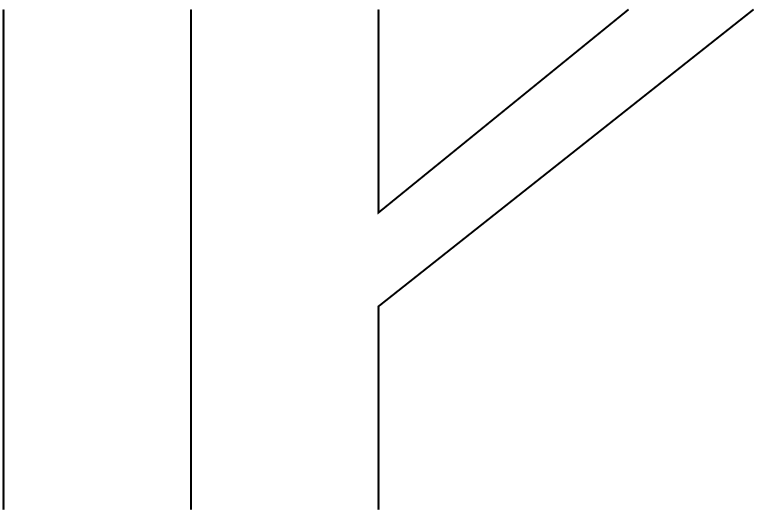,height=5ex,width=3.0em}} \\*[-1ex]
	
 \begin{array}{lcr} 
	{\scriptstyle i}\,&{\scriptstyle j}\,&{\scriptstyle k}
 \end{array} 
 \end{array} 
+ 2 \beta\,
 \begin{array}{r}
 \begin{array}{rrrrr}
	{\scriptstyle q}&{\scriptstyle p}&
	{\scriptstyle l}&{\scriptstyle m}&{\scriptstyle n}
 \end{array} \\*[-.2ex]
	\mbox{\psfig{file=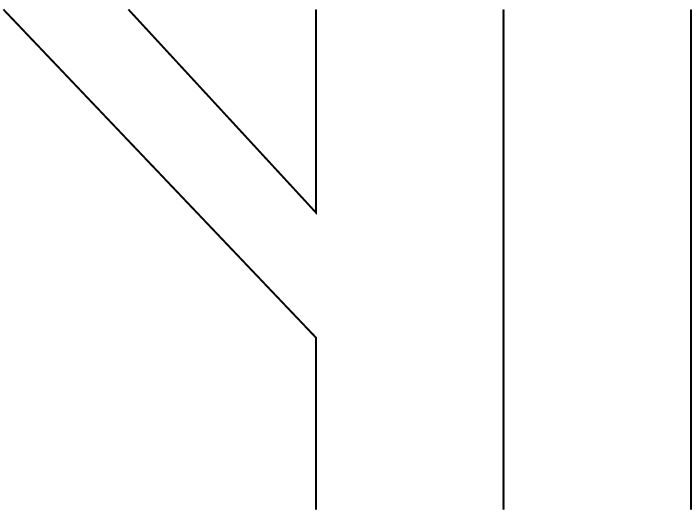,height=5ex,width=3.0em}\ } \\*[-1ex]
 \begin{array}{lcr}
	{\scriptstyle i}\,&{\scriptstyle j}\,&{\scriptstyle k}
 \end{array} 
 \end{array}
+ 2 \gamma\,
 \begin{array}{l}
 \begin{array}{llllr}
	{\scriptstyle l}&{\scriptstyle m}&{\scriptstyle n}&
	{\scriptstyle p}&{\scriptstyle q}
 \end{array} \\*[-.2ex]
	\mbox{\ \psfig{file=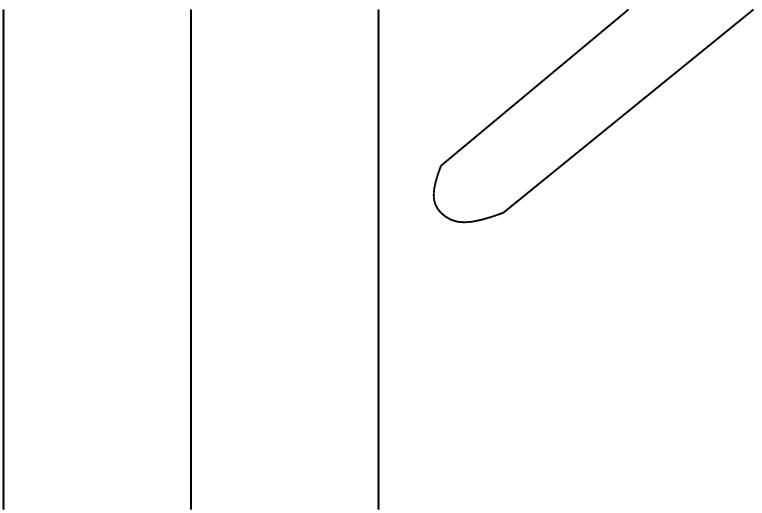,height=5ex,width=3.0em}} \\*[-1ex]
 \begin{array}{lcr} 
	{\scriptstyle i}\,&{\scriptstyle j}\,&{\scriptstyle k}
 \end{array} 
 \end{array}
\right] \,,
\label{eq:BBpi} \\
&\phantom{|}&
 \raisebox{-1.0ex}[4em]{\psfig{file=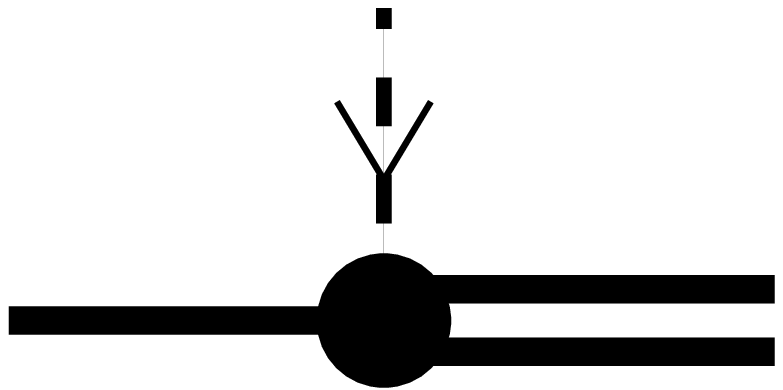,width=7em}} \mu
\ \ = \ \ {k_\mu \over f} \left[ -\sqrt{\scriptstyle 3\over2}\ {\cal C}  \,
\raisebox{-2.5ex}[4.0ex]{\psfig{file=Vertex2.ps,height=6ex,width=3.5em}\ \ } 
\right] \,,
\label{eq:BTpi} \\
&\phantom{|}&
\nu\!\!\!\! \raisebox{-1.0ex}[4em]{\psfig{file=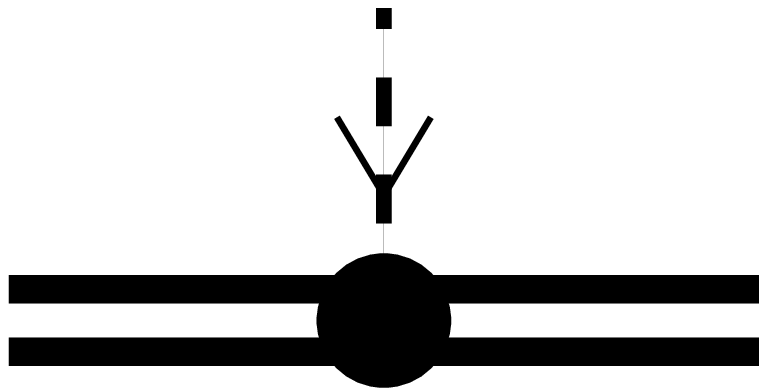,width=7em}} \mu
\ \ =  \ \ {g_{\mu\nu}\, k \cdot S \over f} \left[ \;
2 {\cal H}\ 
\raisebox{-2.5ex}[4.0ex]{\psfig{file=Vertex2.ps,height=6ex,width=3.5em}\ \ }
+ 2 \gamma'\ \ 
\raisebox{-2.5ex}[4.0ex]{\psfig{file=Vertex3.ps,height=6ex,width=3.5em}\ \ }
\right] \,,
\label{eq:TTpi}
\end{eqnarray}
where $k$ is the incoming pion momentum, and $\mu,\nu$ are the Lorentz
indices of the Rarita-Schwinger fields. 
We have shown all flavor indices explicitly for the first vertex
in order to illustrate the conventions that we use in subsequent vertices. 
The two-point vertices are more straightforward
\begin{eqnarray}
\mbox{\psfig{file=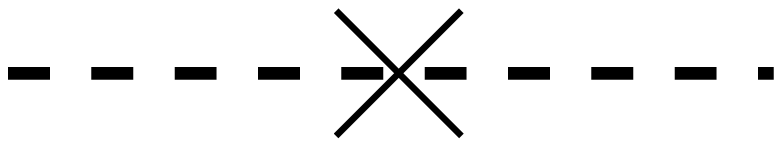,width=7em}}\phantom{\mu}
 &\ =\ & i (\alpha k^2 -m_0^2)\ {2\over3}\ 
\raisebox{-2.5ex}[4.0ex]{\psfig{file=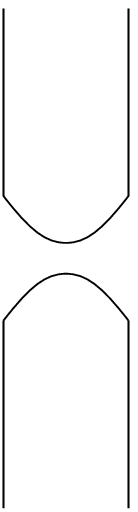,height=7ex}\ \ }
\,,\label{eq:Hairpin}\\
\mbox{\psfig{file=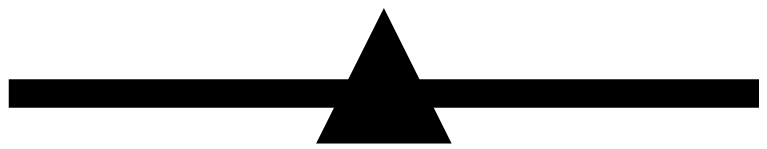,width=7em}} \phantom{\mu}
&\ =\ & i \left[
2 \alpha_m \
\raisebox{-2.5ex}[4.0ex]{\psfig{file=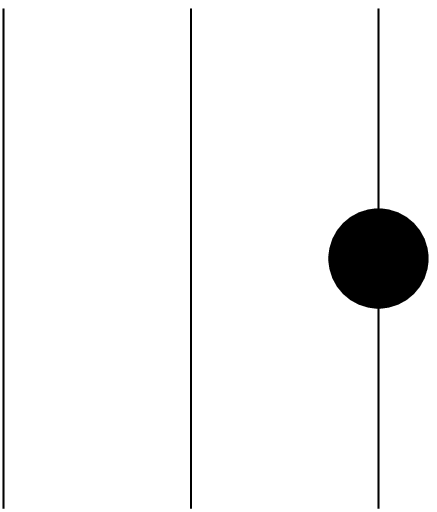,height=6ex}\ \ }
+ 2 \beta_m \
\raisebox{-2.5ex}[4.0ex]{\psfig{file=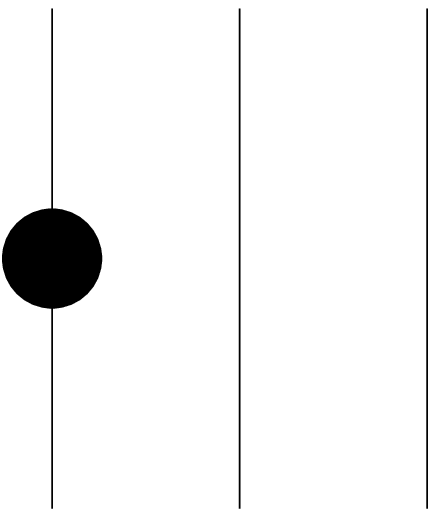,height=6ex}\ \ }
\right]\,,
\label{eq:Bmass}\\
\nu\!\!\!\! \raisebox{-1.0ex}{\psfig{file=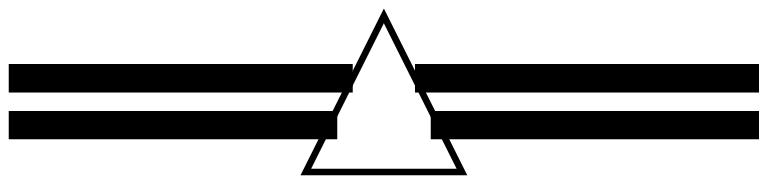,width=7em}} \mu
&\ =\ & i g_{\mu\nu} \Delta M \
\raisebox{-2.5ex}[4.0ex]{\psfig{file=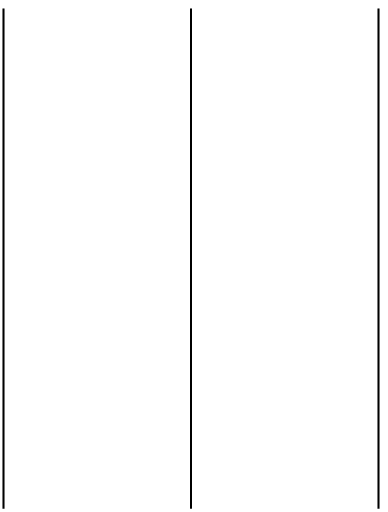,height=6ex}\ \ } \,,
\label{eq:Tdeltam}\\
\nu\!\!\!\! \raisebox{-1.0ex}{\psfig{file=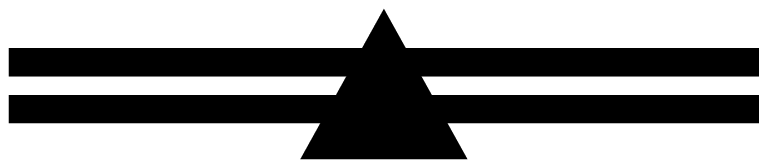,width=7em}} \mu
&\ =\ & i g_{\mu\nu} 2 c \
\raisebox{-2.5ex}[4.0ex]{\psfig{file=Mass1.ps,height=6ex}\ \ } \,,
\label{eq:Tmass}
\end{eqnarray}
where we have adopted the notation that a dot on a line with
flavor $j$ indicates a factor of $m_j$. 
For example, the spin 1/2 mass term is
\[
i\ \delta_{il}\delta_{jm}\delta_{kn}\ (2 \alpha_m m_k + 2 \beta_m m_i) \,.
\]
Note that potential mass terms involving $\sigma$ and $\overline\sigma$
vanish because ${\rm str}(M)=0$.
In the ``hairpin'' vertex, (\ref{eq:Hairpin}), we have included a factor
of 2 resulting from the possible contractions.

We also need vertices in which two pions emanate from a baryon mass term,
but for the sake of brevity we do not give the results here. They are
straightforward extensions of the vertex in Eq.~({\ref{eq:Tmass}).

\subsection{A sample calculation}

To illustrate our method, we sketch the computation of
mass renormalization of the spin-1/2 baryons resulting from the diagrams
of Fig.~(\ref{fig:sunset}). We mainly focus on diagram (a),
and in particular the part proportional to $\beta^2$.
The Feynman integral is
\begin{equation}
	I_{\mu\nu} = \int {d^4k\over(2\pi)^4} 
		{i^2(k_\mu)(-k_\nu)\over (v\!\cdot\! k)(k^2-M_{\pi}^2)} \,,
\end{equation}
and multiplies $\overline u S^\mu S^\nu u$, $u$ being the spinor
of the external state. For the moment we denote the meson mass in the loop
generically as $M_\pi$.
The finite part of the integral is $I_{\mu\nu} = -i g_{\mu\nu} M_\pi/(24\pi)$,
allowing us to eliminate the spin vectors using $\overline u S^2 u = -3/4$.
The mass renormalization, $\Delta m$, is thus given by
\begin{equation}
   -i \Delta m = i\,\beta^2\, {M_{\pi}^3 \over 32\pi f^2} \,,
\label{eq:deltam-deg}
\end{equation}
aside from flavor factors.

\begin{figure}
  \centerline{\psfig{file=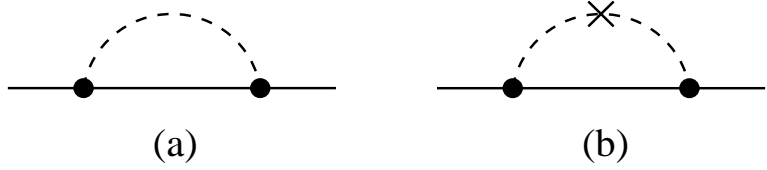,width=4truein}}
\vspace{-0.5truein}
\caption{Examples of diagrams contributing to spin 1/2 baryon mass 
	and wavefunction renormalization.}
\label{fig:sunset}
\end{figure}

The pion in the loop can, for non-degenerate quarks, have one of six masses: 
$M_{ij}^2 = \mu (m_i + m_j)$ where $i\le j\le 3$. 
This is true even if the pion contains bosonic-(anti)quarks,
since they are degenerate with the corresponding quarks.
In our calculation of spin-3/2 baryon mass shifts we have worked with
completely non-degenerate quarks. For the spin-1/2 baryons, on the other hand,
we have considered only the isospin symmetric limit, $\widehat m=m_u=m_d$.
This reduces to three the values of the meson masses:
$M_\pi^2 = 2 \mu \widehat m$, $M_K^2 = \mu(\widehat m + m_s)$ and 
$M_{s \overline s} = 2 \mu m_s$.
The full expression for mass renormalization is then
\begin{equation}
 \Delta m_B = -\,\beta^2
	\sum_{I=\pi,K,s\overline s} c_{B,I} \, {M_I^3 \over 32\pi f^2} \,.
\label{eq:deltam-nondeg} \\
\end{equation}
where $B$ specifies the external spin-1/2 baryon,
and $c_{B,I}$ is the flavor factor.

The flavor factors are obtained by projecting the contribution from
the flavor part of the Feynman rules onto the particular flavor of
the external baryon. Explicitly, we can write
\begin{equation}
	c_{B,I} = \sum_{a,b}
	\psi_{a}^B D^{I}_{ab} \psi_{b}^B \,,
\label{eq:clebsch-formula}
\end{equation}
where the indices $a$ and $b$ run over all 216 values of the 
three flavor indices in the dependent field basis,
$D^I_{ab}$ is the flavor part of the Feynman graph in this basis,
and $B$ is one of the independent baryons, 
whose ``wavefuction'' is $\psi^{B}$. 
The wavefunctions can be read off from
Eqs.\ (\ref{eq:states})-(\ref{eq:tstates}).
The flavor ``charged'' octet baryons are exemplified by
\begin{equation}
   \psi^p = 
\bordermatrix{ & \psi_{112} &\!\!\!\psi_{121} &\!\!\!\psi_{211}\cr
	    \sqrt{\mbox{\small$1\over6$}}\!\! 
		& 1, & 1, & -2 \cr} \,, \qquad
   \psi^{\Xi^-}=\bordermatrix{&\psi_{332}&\!\!\!\psi_{323}&\!\!\!\psi_{233}\cr
	    \sqrt{\mbox{\small$1\over6$}}\!\! 
		& 1, & 1, & -2 \cr} \,,
\label{eq:wf1} 
\end{equation}
where we show only the non-zero elements.
The flavor-neutral octets have wavefuctions
\begin{eqnarray}
 \psi^{\Sigma^0} &=& \bordermatrix{ &
 \psi_{123}&\!\!\!\psi_{231}&\!\!\!\psi_{312}&\!\!\!
 \psi_{132}&\!\!\!\psi_{213}&\!\!\!\psi_{321}\cr
	\sqrt{\mbox{\small$1\over12$}}\!\! 
	& 1, & 1, & -2, & 1, & 1, & -2 \cr} \,,
\label{eq:wf2} \\*[+1ex]
 \psi^{\Lambda} &=& \bordermatrix{ &
 \psi_{123}&\!\!\!\psi_{231}&\!\!\!
 \psi_{132}&\!\!\!\psi_{213} \cr
\mbox{\small$1\over2$} & 1, & -1, & 1, & -1 \cr}\,.
\label{eq:wf3}
\end{eqnarray}
In other calculations, we also need the wavefuctions for the decuplet
baryons, $\psi^T_a$. These are exemplified by
\begin{eqnarray}
   \psi^{\Delta^{++}} &=& 
\bordermatrix{ & \psi_{111}\cr
	       & 1 \cr} \,, \qquad
\psi^{\Sigma^{*,+}} =
\bordermatrix{ & \psi_{113} &\!\!\!\psi_{131} &\!\!\!\psi_{311}\cr
	    \sqrt{\mbox{\small$1\over3$}}\!\! & 1, & 1, & 1 \cr} \,, 
\label{eq:wf4} \\*[+1ex]
\psi^{\Sigma^{*,0}} &=&
\bordermatrix{&	\psi_{123}&\!\!\!\psi_{231}&\!\!\!\psi_{312}&\!\!\!
		\psi_{132}&\!\!\!\psi_{213}&\!\!\!\psi_{321}\cr
	    \sqrt{\mbox{\small$1\over6$}}\!\! 
		& 1, & 1, & 1, & 1, & 1, & 1 \cr} \,.
\label{eq:wf5} 
\end{eqnarray}
Explicit forms for the wavefunctions involving bosonic-quarks are not needed,
since these do not appear as external states in quenched lattice calculations.

The wavefuctions satisfy various useful properties.
Those for the spin-1/2 and spin-3/2 baryons are separately othonormal
\begin{equation}
\sum_{a} \psi^{B}_{a} \psi_{a}^{B'} = 
\left\{ \begin{array}{ll}
		1 & B=B'\cr
	    	0 & B\ne B' 
	\end{array} \right.
\,,\qquad
\sum_{a} \psi^{T}_{a} \psi_{a}^{T'} = 
\left\{ \begin{array}{ll}
		1 & T=T'\cr
	    	0 & T\ne T' 
	\end{array} \right.
\,.
\end{equation}
This ensures that the kinetic terms are correctly normalized when
written in terms of independent fields.
The wavefuctions are eigenvectors of the flavor propagators
with unit eigenvalues
\begin{equation}
{\cal F}^{(1/2)}_{ab}\, \psi^B_b = \psi^B_a\,,\qquad
{\cal F}^{(3/2)}_{ab}\, \psi^T_b = \psi^T_a\,.
\end{equation}
All other eigenvectors have zero eigenvalues, corresponding to the
fact that the flavor propagators are projection matrices,
i.e. $[{\cal F}^{(j)}]^2={\cal F}^{(j)}$ for $j=1/2$ and $3/2$.
Thus the flavor propagators can be written as
\begin{equation}
{\cal F}^{(1/2)}_{ab} = \sum_B \psi^B_a \psi^B_b \,,\quad
{\cal F}^{(3/2)}_{ab} = \sum_T \psi^T_a \psi^T_b \,.
\end{equation}
The sums here are over the independent baryons of the appropriate spin
(70 and 38 respectively), including baryons containing bosonic-quarks.
These results show that the baryon propagators in Feynman diagrams 
project onto the independent states, as claimed above.

Returning to the sample calculation, we next compute 
the ``matrix elements'' $D^{I}_{ab}$ using the flavor Feynman rules.
We begin with the elements needed if the external baryon is a proton,
i.e. $a,b=112,121,211$.
A representative element is
\begin{eqnarray}
D^{\pi}_{112,112}  & &
   \begin{array}[t]{cccccl}
	 \;=\;& \phantom{\displaystyle{1\over6}} 
		\left(\;{\displaystyle \sum_{l=\{1,2,4,5\}}}
		\quarkflow{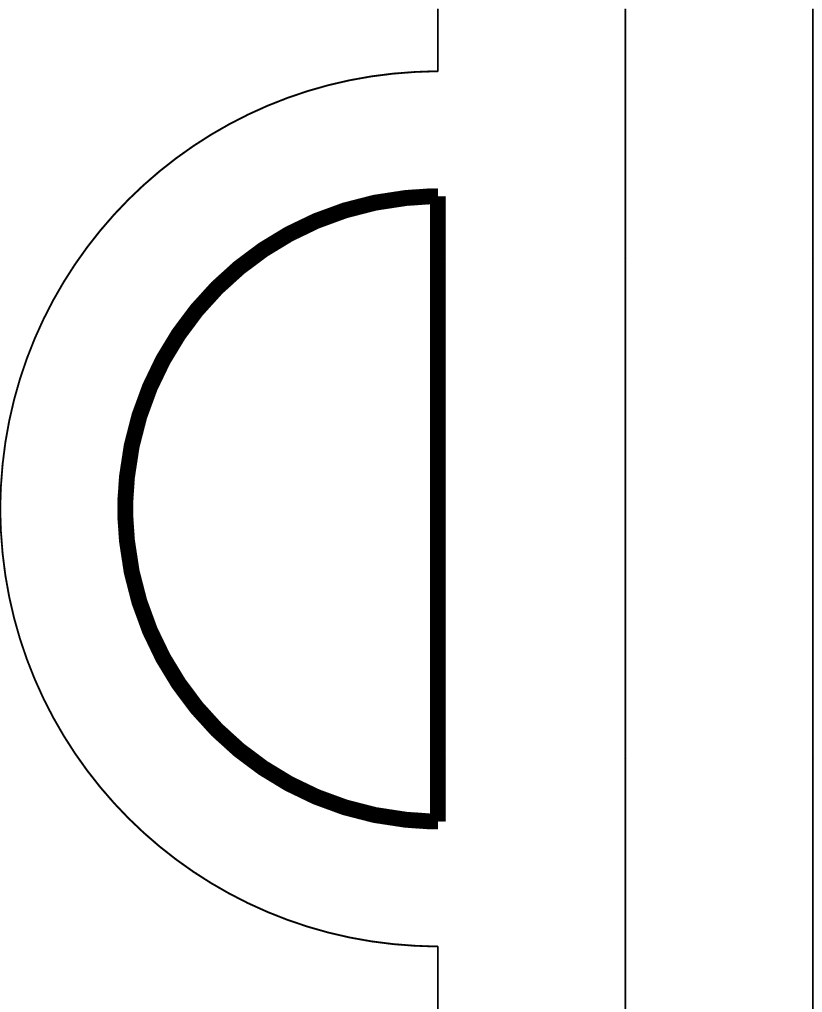}{1}{1}{2}\;\right)
		\hspace{-2.9em}{\scriptstyle l}\hspace{+2.9em}
	&\;+\;& \quarkflow{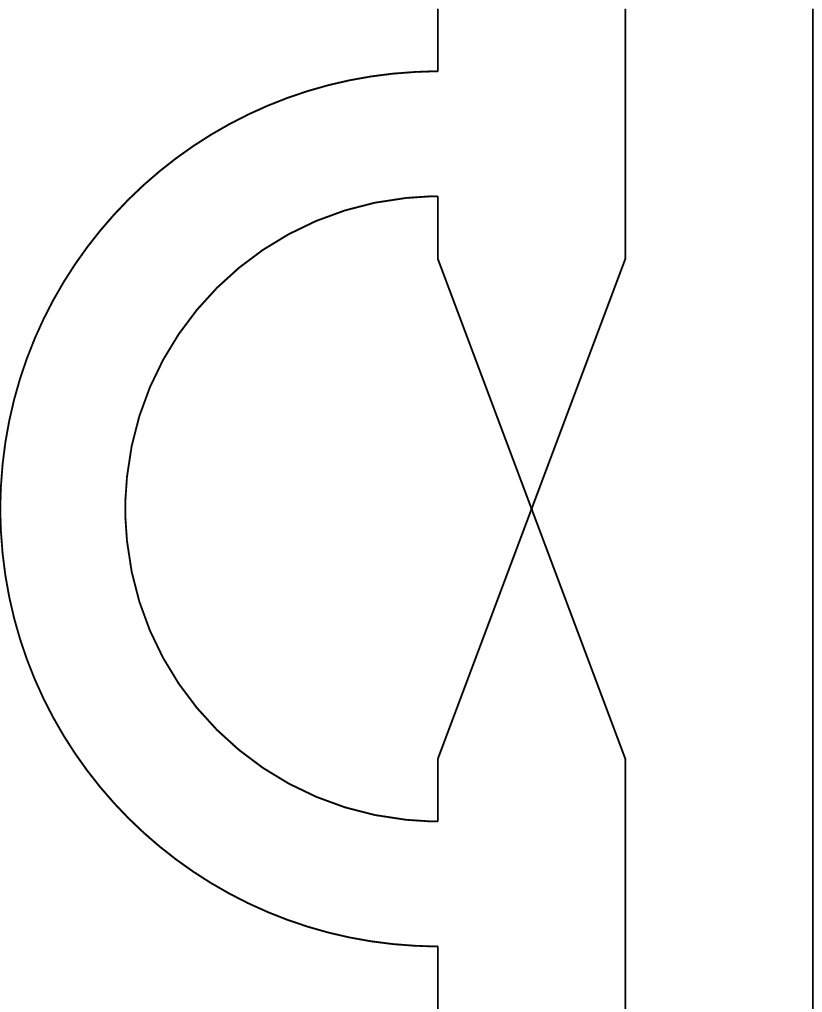}{1}{1}{2}
	&\;+\;& \quarkflow{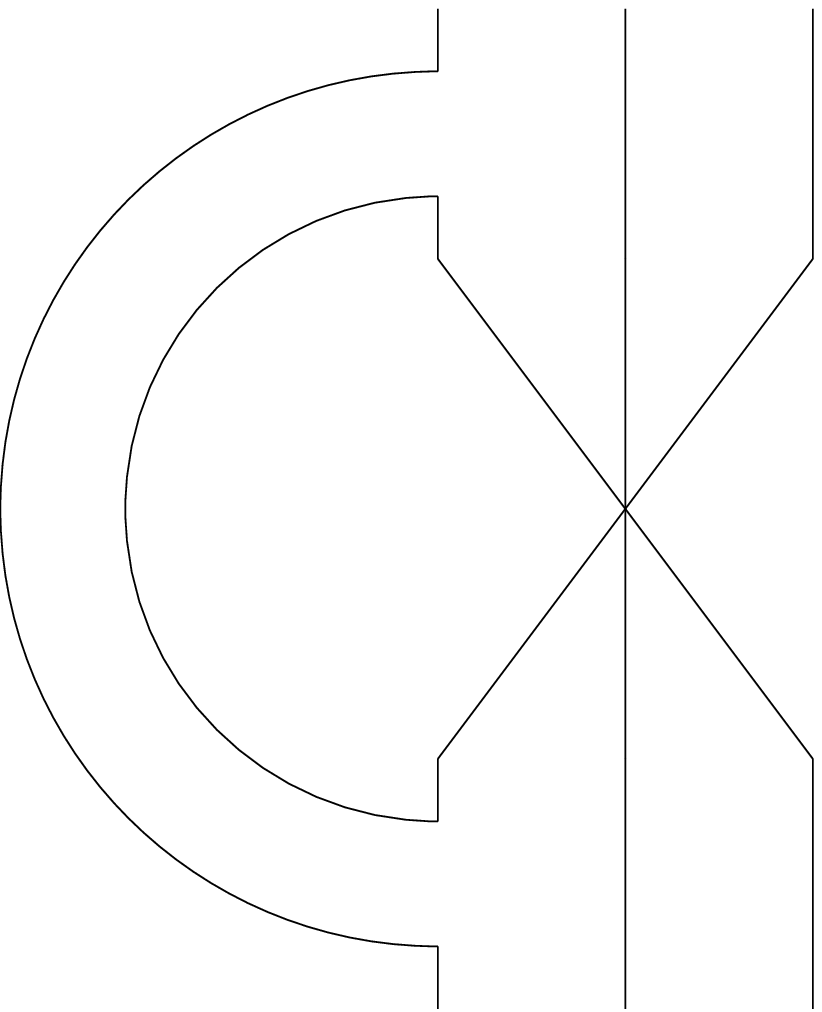}{1}{1}{2}
    \\*[+5ex]
	 \;=\;& {\displaystyle{1\over6}} 
		\left(\;{\displaystyle \sum_{l=\{1,1,0,0\}}}
		(-1)^{l+1}\right)
	&\;-\;& {\displaystyle{1\over12}}
	&\;-\;& \hspace{1.5em} {\displaystyle{1\over12}} 
		\quad \;=\; \quad {\displaystyle{-{1\over6}}},
   \end{array}
\label{eq:Dpi112} \\*[+5ex]
D^{K}_{112,112}  & &
   \begin{array}[t]{cl}
	\;=\;& \phantom{\displaystyle{1\over6}} 
		{\displaystyle \sum_{l=\{3,6\}}}
		\quarkflow{BetaBeta1.ps}{1}{1}{2}
		\hspace{-1.8em}{\scriptstyle l}\hspace{+1.8em}
	  \\*[+5ex]
	\;=\;& {\displaystyle{1\over6}} 
		{\displaystyle \sum_{l=\{1,0\}}}
		(-1)^{l+1} \quad \;=\; \quad 0,
   \end{array}
\label{eq:DK112} \\*[+5ex]
D^{s\overline s}_{112,112} & & 
	\begin{array}[t]{cl}
		\;=\;& 0 
	\end{array}
\label{eq:Dss112}
\end{eqnarray}
In these equations we are using the diagrammatic notation in a slightly
different way from above---the diagrams denote 
both the way in which the flavor indices are contracted
and the corresponding factors from the propagators and vertices.
The remaining elements can be evaluated similarly, with the results
\begin{equation}
   D^{\pi} \;=\; -{1\over6}
	\left( \begin{array}{rrr}	
			1 & 1 & 0 \\
			1 & 1 & 0 \\
			0 & 0 & 2
	\end{array} \right)\,, \qquad 
   D^{K} \;=\; 0\,,	\qquad 
   D^{s\overline s} \;=\; 0\,,
\end{equation}
for the $(112,121,211)$ block.
Sandwiching this between the proton wavefunctions gives the flavor
factors $c_{p,I}=-1/3,0,0$, for $I=\pi,K,{s\overline s}$.
These can be inserted in Eq.~(\ref{eq:deltam-nondeg}) to obtain
the proton mass renormalization.

It is worthwhile contrasting the calculation thus far with that in QCD.
To make the comparison, it is better to use
a three index field to represent spin-1/2 baryons in QCD,
for then one can develop the calculation diagrammatically as above. 
The major difference from QQCD is that the indices run from 1-3 instead of 1-6.
(One must also project against the $\eta'$ in the meson propagator in QCD,
but this is a small effect numerically.)
Thus the main difference between calculations in the two theories is that 
quark-flow diagrams involving internal loops cancel in QQCD, 
but do not in QCD.
This is, qualitatively, exactly as expected.
What we are able to do here
is make a quantitative calculation of the effect of this cancelation.

A striking feature of the calculation for baryons is that 
there are quark-flow diagrams which contribute in QCD which
survive the cancelation between $q$ and $\widetilde q$ loops,
e.g. the last two diagrams contributing to $D^\pi_{112,112}$.
This is in contrast to mesonic quantities (e.g. $M_\pi$ and $f_\pi$), 
where the cancelation of quark-loops removes all the QCD contributions
\cite{jm,sharpe1}.
The surviving diagrams for baryons must, however, contain mesons
composed of quarks having the flavors of the valence quarks of the
external baryon. 
Thus, since the proton does not contain a valence strange quark, there can
be no contributions from kaons or $s\overline s$ mesons in the loop.
If the external particle contains a strange quark, however, kaon 
contributions are present, e.g.
\begin{displaymath}
\begin{array}{lcllcl}
	D^{\pi}_{113,113}
	&\;=\;& \quarkflow{BetaBeta3.ps}{1}{1}{3} 
		\;=\; {\displaystyle{-{1\over12}}}\,,
	\qquad &
	D^{K}_{113,113}
	&\;=\;& \quarkflow{BetaBeta6.ps}{1}{1}{3} 
		\;=\; {\displaystyle{-{1\over12}}}\,,
\\*[+6ex]
	D^{\pi}_{131,113}
	&\;=\;& \quarkflow{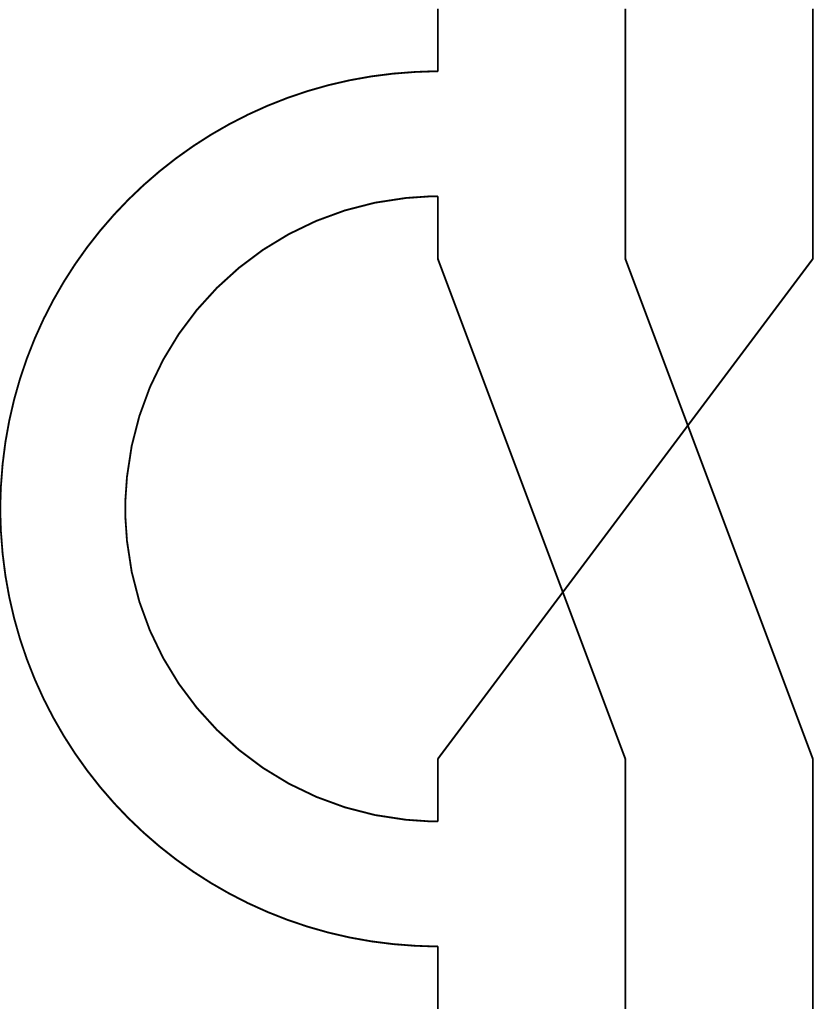}{1}{1}{3} 
		\;=\; {\displaystyle{-{1\over12}}}\,
	\qquad &
	D^{K}_{131,113}
	&\;=\;& \quarkflow{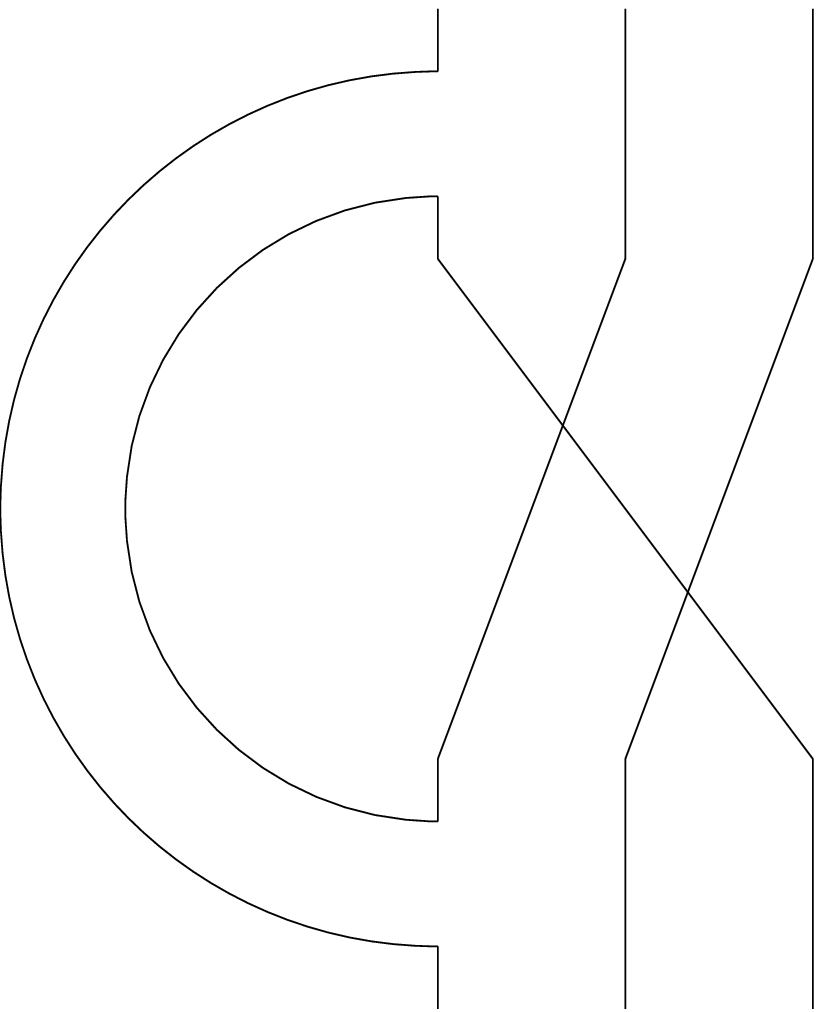}{1}{1}{3} 
		\;=\; {\displaystyle{-{1\over12}}}\,.
\end{array}
\end{displaymath}
\\
Here we have not shown the quark-loop cancellations explicitly.

Proceeding as above, we calculate all the relevant parts of $D^{I}$.
Contracting these with the wavefunctions, we obtain the flavor factors
\begin{equation}
	- 18\, c_{B,I} \;=\; \bordermatrix{
	    & \pi & K & s\overline s \cr
	 p  & 6 & 0 & 0 \cr 
     \Sigma & 1 & 5 & 0 \cr
	\Xi & 0 & 5 & 1 \cr 
    \Lambda & 3 & 3 & 0 \cr}\,.
\label{eq:clebschs}
\end{equation}
When inserted in Eq.~(\ref{eq:deltam-nondeg}) this gives the result
for the contribution proportional to $\beta^2$
coming from Fig.~(\ref{fig:sunset}a). 

The calculation of the other contributions proceeds similarly.
The diagrams for the $\alpha^2$ terms are the same as those just discussed,
except that the loops come off the right hand side of the baryon instead
of the left. This changes the flavor factors.
The diagrams for the $\alpha\beta$ contributions are different, and are
shown in Fig. \ref{fig:sunset2}. Again, some of the diagrams
involve internal quark loops and cancel in the quenched approximation,
whereas others remain.
Diagrams for the $\gamma\alpha$ contributions are shown in 
Fig.~\ref{fig:sunset3}. These are special to the quenched approximation
because only flavor singlet pions can appear in the loop.
The same is true of contributions proportional to $\gamma\beta$.
Finally, terms proportional to $\gamma^2$ vanish because they involve
internal quark loops.

\begin{figure}
  \centerline{\psfig{file=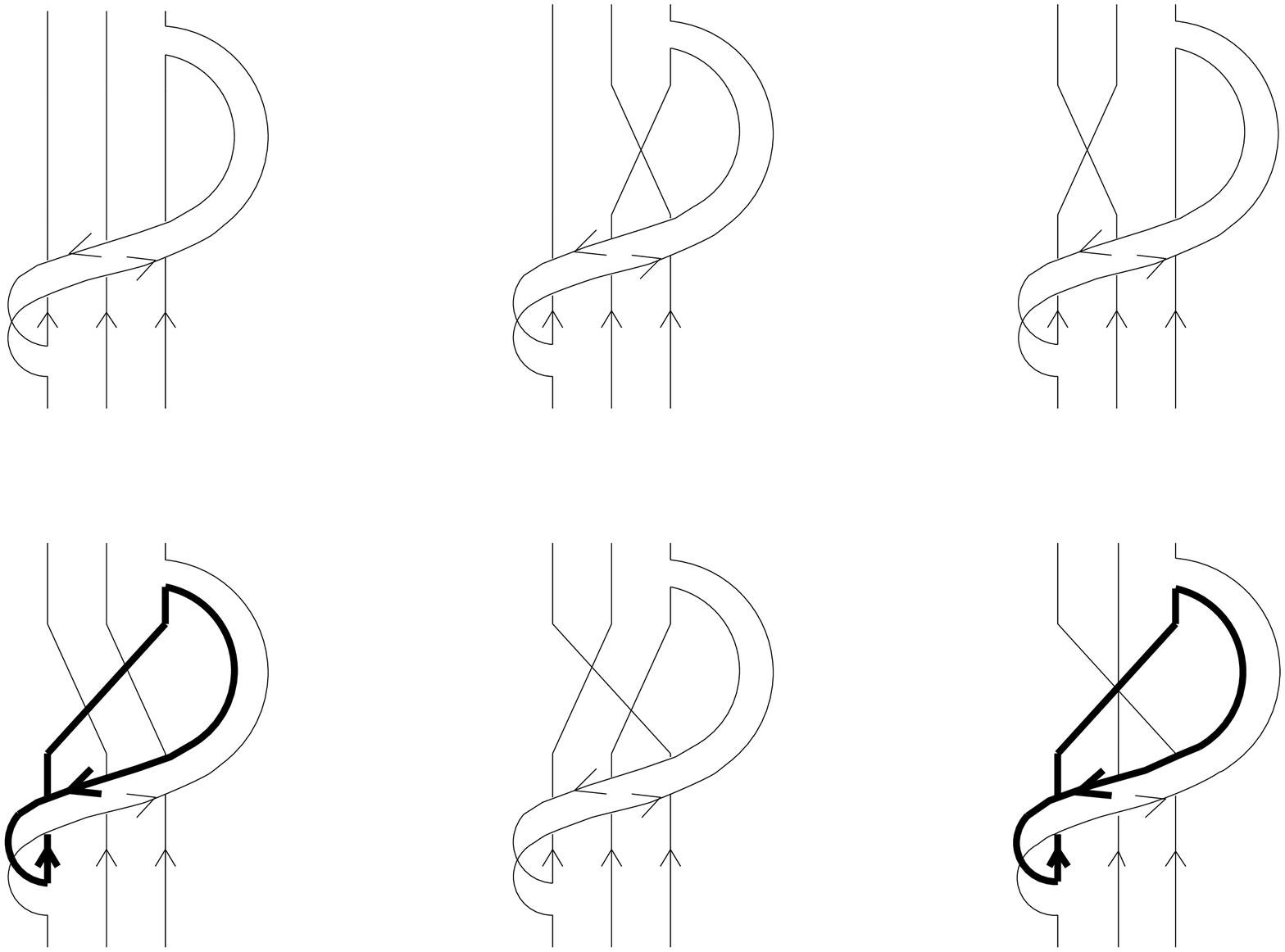,height=6truecm,width=12truecm}}
  \caption{Quark-flow diagrams contributing to the $\alpha\beta$ terms in
	mass renormalization. Arrows have been added to help clarify 
	quark flows. The boldface line can be both a $q$ and a $\widetilde q$;
        other lines represent only quarks.}
\label{fig:sunset2}
\end{figure}

\begin{figure}
 \centerline{\psfig{file=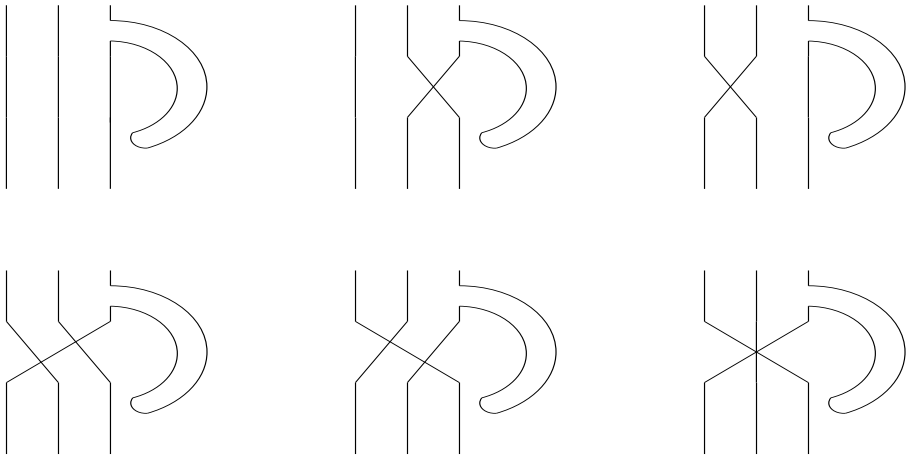,height=6truecm,width=12truecm}}
  \caption{Quark-flow diagrams contributing to the $\gamma\alpha$ terms in
	mass renormalization.}
\label{fig:sunset3}
\end{figure}

Additional types of quark flow occur in the evaluation
of Fig.~(\ref{fig:sunset}b). This diagram is not present in QCD
because only flavor singlet pions couple to the 
$m_0^2$ (or $\alpha_\Phi$) vertex.
Examples of the corresponding quark-flow diagrams, 
for contributions proportional to $\beta^2 m_0^2$, 
are shown in Fig.~\ref{fig:sunset4}. 
There are similar diagrams for the $\alpha^2$ and $\alpha\beta$ contributions.
It is these type of diagrams which are the sole quenched contribution
to mesonic quantities.
Finally, contributions proportional to $\gamma$ involve quark loops 
and cancel.

\begin{figure}
 \centerline{\psfig{file=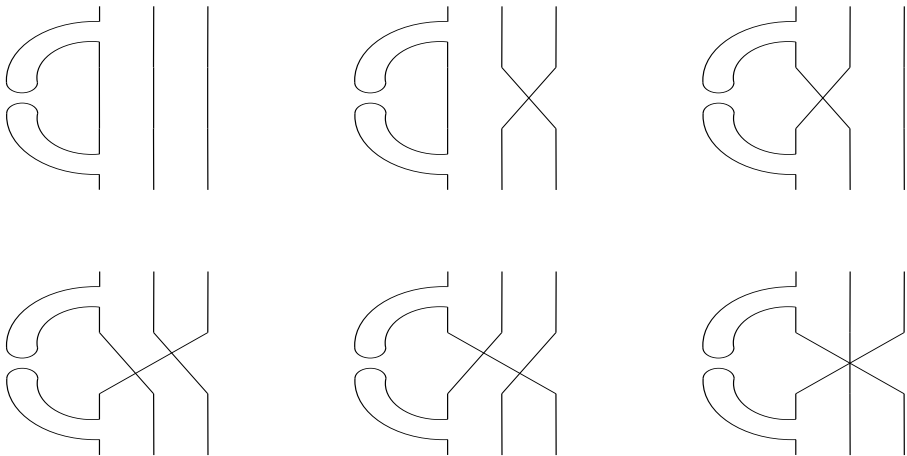,height=6truecm,width=12truecm}}
  \caption{Quark-flow diagrams contributing to the $\beta^2 m_0^2$ terms in
	mass renormalization.}
\label{fig:sunset4}
\end{figure}

\subsection{Quenched oddities}
\label{sec:deltadecay}

We close this section by noting some peculiar features of the workings
of chiral perturbation theory for QQCD.
We first show why care must be taken if one 
attempts to calculate quenched results
without the benefit of the graded formalism.
This might be done, for example,
by making a quark-flow correspondence using the QCD chiral Lagrangian.
Such an approach was used successfully for mesons in Ref. \cite{sharpe1}.

Consider $D^\pi_{211,211}$, which can be split up as follows
\begin{eqnarray}
D^{p,\pi}_{211,211}
	&=& \quarkflow{BetaBeta1.ps}{2}{1}{1}
                \hspace{-1.7em}{\scriptstyle 1}\hspace{+1.7em}
        \;+\; \quarkflow{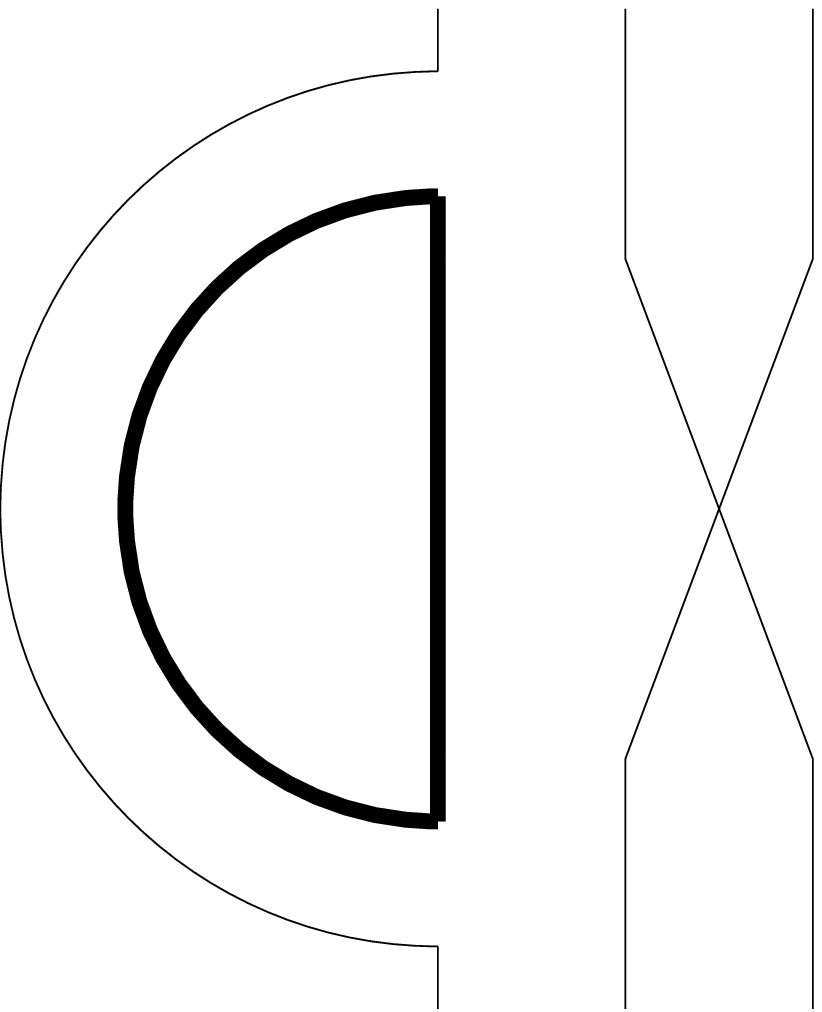}{2}{1}{1}
                \hspace{-1.7em}{\scriptstyle 1}\hspace{+1.7em}
        \;+\; \quarkflow{BetaBeta3.ps}{2}{1}{1}
        \;+\; \quarkflow{BetaBeta4.ps}{2}{1}{1}
        \;+\; \quarkflow{BetaBeta5.ps}{2}{1}{1}
        \;+\; \quarkflow{BetaBeta6.ps}{2}{1}{1} \;+\;
\nonumber\\*[+\belowdisplayskip]
	&+& {\displaystyle \sum_{l=\{2,4,5\}}}\left(
	    \quarkflow{BetaBeta1.ps}{2}{1}{1}
               \hspace{-1.7em}{\scriptstyle l}\hspace{+1.7em}
	\;+\; \quarkflow{BetaBeta2.ps}{2}{1}{1}
               \hspace{-1.7em}{\scriptstyle l}\hspace{+1.7em}\;\right).
\end{eqnarray}
The correct approach is to drop all diagrams containing quark loops,
since these cancel in pairs.
This leaves the last four diagrams on the first line, which give
a non-zero result, $-1/3$.

One might have been tempted, however, to argue as follows.
In each of these four diagrams, the baryon propagating
in the loop is ${\cal B}_{111}$. 
There is, however, no spin-1/2 baryon consisting of three up quarks---indeed, 
the symmetry relations set ${\cal B}_{111}=0$.
Thus $D^\pi_{211,211}$ should vanish in the quenched approximation.

This argument is wrong because, as we have seen, the baryon propagator
automatically projects against ${\cal B}_{111}$.
What happens is that the {\em six} diagrams on the first line 
(the only ones containing ${\cal B}_{111}$) cancel: 
their sum is proportional to $(2+2-1-1-1-1)=0$.
In the second line, the contributions from $l=2$ and $l=5$ cancel
as usual, leaving the entire contribution to come from the diagram with $l=4$. 
This consists entirely of diagrams containing an internal $\widetilde u$ loop! 
Thus, if one insists on removing ${\cal B}_{111}$, then to obtain
the correct result in the quenched theory one must include bosonic quarks.
Clearly the simplest approach is to let the theory take care of
doing the projections itself.

\begin{figure}
\centerline{\psfig{file=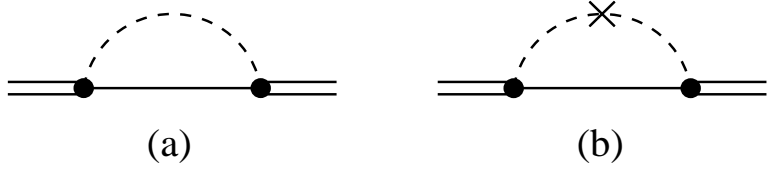,width=4truein}}
\vspace{-0.5truein}
\caption{Diagrams contributing to the $\Delta$ self-energy which can
lead to $\Delta$ decay.}
\label{fig:deltadecay}
\end{figure}

A related peculiarity concerns the decay of decuplet baryons in QQCD.
In QCD they decay through the strong interactions,
e.g. $\Delta^{++} \to \pi^+ p$,
due to the vertex proportional to ${\cal C}$.
One could study this numerically by calculating the
$\Delta$ two-point function in Euclidean space.
At large Euclidean times it would be dominated by the $\pi^+ p$
intermediate state, resulting from the cut in Fig.~(\ref{fig:deltadecay}a).%
\footnote{%
Strictly speaking, this is only true in large enough volumes. 
The decay amplitude is p-wave, and so vanishes at threshold,
requiring the intermediate pion to have non-zero momentum. 
This increases the threshold for the decay on small lattices.}
The issue is whether the same is true in QQCD.
The following argument suggests that it is not.
The $\Delta^{++}$ has quark composition $uuu$, and, since there
are no quark loops in QQCD, it can decay only into a $u \overline{u}$ meson
and a $uuu$ spin-1/2 baryon. But there is no $uuu$ octet baryon, 
and so the decay cannot occur.
By isospin symmetry, all of the $\Delta$'s must be stable in QQCD.

This argument is wrong for the same reason that $D^\pi_{211,211}$ does
not vanish. One can only impose the requirement that the ${\cal B}_{111}$
does not contribute {\em if one works with the bosonic-quark formalism
in which all symmetries are manifest}. But then there are additional
contributions to the self-energy which do lead the $\Delta$ to decay.
If one instead simply throws away all quark loops, {\em then 
the $uuu$ baryon does contribute in intermediate states}, 
and again the $\Delta$ decays.
Another way of understanding this is to note that quark loops 
are necessary in order to implement the Pauli exclusion principle 
on internal propagators.

Thus we claim that one can study the decay of the $\Delta$ even in QQCD.
This is such a counter-intuitive result that it is worthwhile understanding
in more detail. Consider the quark flow diagrams contributing to the
${\cal C}^2$ term in Fig.~(\ref{fig:deltadecay}a)
\smallskip
\begin{eqnarray}
D^{\pi}_{111,111}
	&=& \quarkflow{BetaBeta1.ps}{1}{1}{1}
                \hspace{-1.7em}{\scriptstyle 1}\hspace{+1.7em}
        \;+\; \quarkflow{BetaBeta2.ps}{1}{1}{1}
                \hspace{-1.7em}{\scriptstyle 1}\hspace{+1.7em}
        \;+\; \quarkflow{BetaBeta3.ps}{1}{1}{1}
        \;+\; \quarkflow{BetaBeta4.ps}{1}{1}{1}
        \;+\; \quarkflow{BetaBeta5.ps}{1}{1}{1}
        \;+\; \quarkflow{BetaBeta6.ps}{1}{1}{1} \;+\;
\nonumber\\*[+\belowdisplayskip]
	&+& {\displaystyle \sum_{l=\{2,4,5\}}}\left(
	    \quarkflow{BetaBeta1.ps}{1}{1}{1}
               \hspace{-1.7em}{\scriptstyle l}\hspace{+1.7em}
	\;+\; \quarkflow{BetaBeta2.ps}{1}{1}{1}
               \hspace{-1.7em}{\scriptstyle l}\hspace{+1.7em}\;\right)\,.
\end{eqnarray}
\smallskip\noindent
The quark-flows are the same as those contributing 
to the $\beta^2$ part of $D^{\pi}_{211,211}$---the only difference
is the external flavors.
The story is now the same as above. The diagrams with quark loops
cancel, leaving a non-zero contribution from the last four diagrams
on the first line---all of which involve ${\cal B}_{111}$.
Alternatively, the Pauli exclusion principle requires that the diagrams
on the first line cancel, leaving a non-zero contribution from the
diagrams of the second line with $\ell=4$. In this view, the
quenched decay is actually $\Delta^{++} \to (u\overline{\widetilde u}) +
(\widetilde u u u)$.
In QCD, by contrast, the diagrams on the first line still cancel,
but the non-vanishing contribution is from $\ell=2$ on the second line.
Thus the QCD and QQCD decay amplitudes have the same
magnitude but opposite sign (assuming $m_u=m_d$).

Actually this is not the whole story in QQCD. There are 
contributions from Fig.~(\ref{fig:deltadecay}a) 
proportional to ${\cal C}\gamma'$, and there are contributions
from Fig.~(\ref{fig:deltadecay}b) proportional to ${\cal C}^2 m_0^2$
and ${\cal C}^2 \alpha_\Phi^2$.
Both give rise to a pion-nucleon cut in the $\Delta$ propagator.
Thus the $\Delta$ will have a different width in QQCD than in QCD.
Nevertheless, it will be interesting to use QQCD as a testing ground
for methods to study an unstable $\Delta$.
This requires, of course, that $m_\Delta - m_p > m_\pi$,
a condition which is not close to being satisfied in most present simulations.
One must use lighter quarks (to increase $m_\Delta -m_p$ and to
decrease $m_\pi$) and larger lattices (to decrease the cost of having
non-zero momentum pions) for the decay to be kinematically allowed.

\section{Results}
\label{sec:4}

Using the methods explained above, 
we have calculated mass renormalization of octet and decuplet
baryons in QQCD resulting from the diagrams shown in Fig.~\ref{fig:feynman}.
These diagrams lead to the dominant corrections in a limit which
we explain below.
The calculation of the flavor factors is quite laborious. 
In order to check our results, we have done two independent calculations, 
one by hand and the other using Mathematica.
We have also tested our method by repeating the calculation for QCD,
and comparing to existing results.

We present our results separately for each subset of graphs,
i.e. $(a)-(e)$ in Fig. \ref{fig:feynman}.
We use the notation
\begin{equation}
M_{\rm bary}(m_u,m_d,m_s) = M_0 + 
\sum_{G=a,b,c,d,e} M_{\rm bary}^{(G)}(m_u,m_d,m_s) \,.
\end{equation}
Here $M_0$ is the octet baryon mass in the chiral limit.
For the octet baryons we have done the calculation assuming $m_u=m_d$,
while for the decuplets we have allowed all three quarks to have
different masses.

\begin{figure}[tb]
  \centerline{\psfig{file=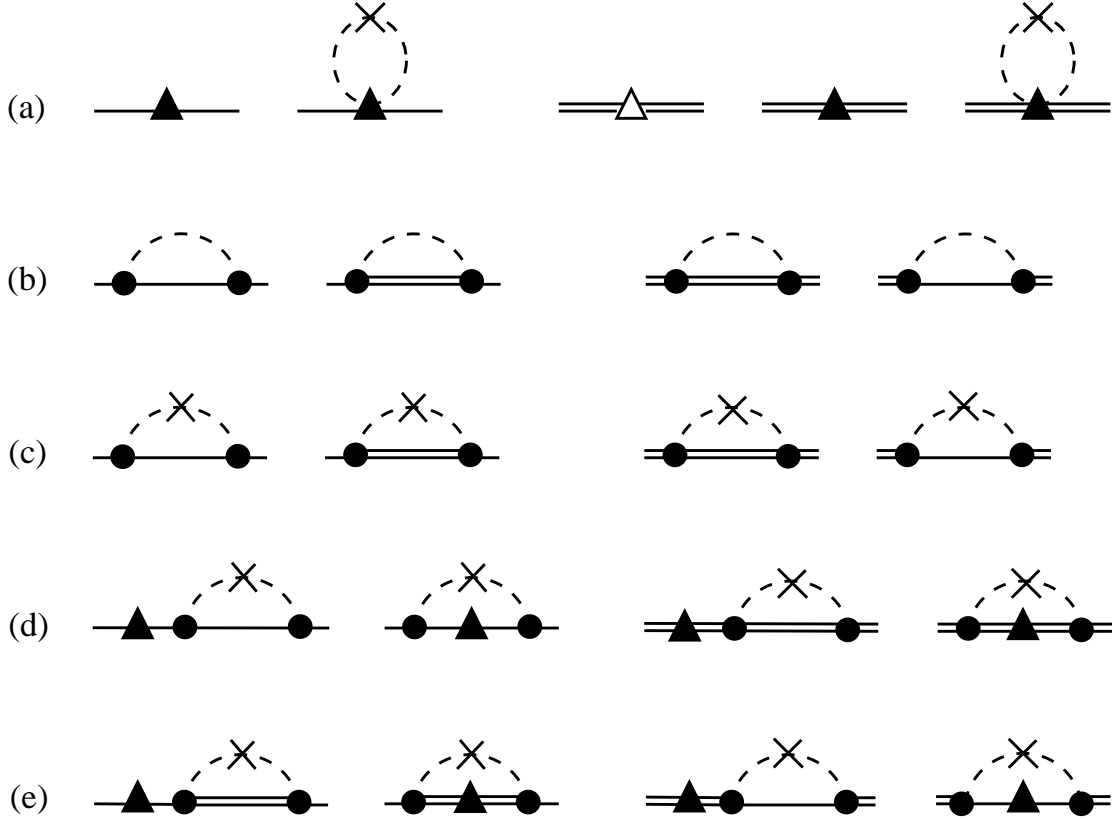,width=5.8truein}}
  \caption{Diagrams included in this calculation.
        Those in which a loop is attached to an external line [e.g. the
	first and third diagrams in (d)] represent mass shifts
	coming from wavefunction renormalization.}
\label{fig:feynman}
\end{figure}

A considerable simplification results from the fact that,
in QQCD, the mass of a given baryon depends only on the
masses of the quarks that it contains.
This means that we need only give results for the
$\Sigma$ and $\Lambda$ octets, 
and the $\Sigma^{*,0}$ (which has composition $uds$) decuplet.
Mass shifts for the other baryons
can be found simply by changing the quark masses, e.g.
\begin{equation}
\begin{array}{lll}
M_\Sigma = M_\Sigma(m_u,m_s) \,,\quad
&M_p = M_\Sigma(m_u,m_u) \,,\quad
&M_\Xi = M_\Sigma(m_s,m_u) \,,\quad \\
M_{\Sigma^{*,0}} = M_{\Sigma^*}(m_u,m_d,m_s) \,,
&M_\Delta^{++} = M_{\Sigma^*}(m_u,m_u,m_u) \,,
&M_\Omega^{-} = M_{\Sigma^*}(m_s,m_s,m_s) \,.
\end{array}
\end{equation}
An important check on our results is that all octet baryons are
degenerate if the quarks are degenerate, which implies 
$M_\Sigma(m_u,m_u) = M_\Lambda(m_u,m_u)$.

The diagrams of Fig,~(\ref{fig:feynman}a) give contributions 
proportional to $M_\pi^2$
\begin{eqnarray}
M_\Sigma^{(a)} &=&  -4 b_F M_{uu}^2 + 2(b_D-b_F) M_{ss}^2\,, 
\\
M_\Lambda^{(a)} &=& 4 (2b_D/3-b_F) M_{uu}^2 - 2(b_D/3+b_F) M_{ss}^2\,, 
\\
M_{\Sigma^*}^{(a)} &=& \Delta M + c'(M_{uu}^2+M_{dd}^2+M_{ss}^2) \,,
\end{eqnarray}
where $c'= c/(3\mu)$.
These results require some explanation. 
\begin{enumerate}
\item
The diagrams with filled triangles give contributions 
proportional to quark masses.
These corrections are the same as in QCD, except that 
``sigma-term'' contributions proportional to $\sigma$ or 
$\overline\sigma$ are absent in QQCD.
To make the comparison with QCD more direct,
we have rewritten the octet results using Eq.~(\ref{eq:Qparameters})
to express $\alpha_M$ and $\beta_M$ in terms of $b_D$ and $b_F$.
\item
The decuplet masses are also shifted by $\Delta M$,
the octet-decuplet mass splitting in the chiral limit.
\item
The diagrams involving meson loops with the hairpin vertex
give corrections proportional to $m_q m_0^2 \log(m_\pi^2)$,
which are quenched artifacts.
For small enough $m_\pi$, these
can become larger than the analytic corrections proportional to $m_q$.
They are of the same form as the leading chiral correction to the
pion mass \cite{bg,sharpe1}
\begin{equation}
M_{qq}^2 = 2 \mu m_q 
	\left[ 1 - \delta \log{M_{qq}^2 \over \Lambda_{\chi}^2}\right] \,,
\label{eq:1looppion}
\end{equation}
where\footnote{%
Strictly speaking the meson mass in the logarithm should be the lowest order
approximation ($2\mu m_q$), but it can be replaced by the actual mass
to the order we are working.}
$\delta= {m_{0}^2/48\pi^2 f^2}$.
It turns out that the corrections to baryon masses from these loop graphs 
are of exactly the form to allow them to be absorbed by
replacing quark masses with the corresponding 1-loop corrected
meson squared masses of Eq.~(\ref{eq:1looppion}).
This is convenient, since it is more straightforward to extract
meson masses than quark masses from simulations. 
\end{enumerate}

The diagrams of Fig.~(\ref{fig:feynman}b) are present in both QCD and QQCD.
As shown in the example of Sec.~(\ref{sec:3}), these give rise to
non-analytic terms proportional to $M_\pi^3$.
We find
\begin{eqnarray}
M_\Sigma^{(b)} &=&  
(\mbox{\small$2\over3$} D^2 - 2F^2 -4 F \gamma  
- \mbox{\small$1\over9$} {\cal C}^2) V_{uu}
\nonumber\\
&&\mbox{} 
+(\mbox{\small$2\over3$} D^2 - 4DF + 2F^2 
- \mbox{\small$5\over9$} {\cal C}^2) V_{us} +2 (D-F) \gamma V_{ss}
\\
M_\Lambda^{(b)} &=& 
\left[\mbox{\small$2\over9$} D^2 
- \mbox{\small$8\over3$} D F + 2 F^2 + 4 (2D/3-F) \gamma 
- \mbox{\small$1\over3$} {\cal C}^2\right]\, V_{uu}
\nonumber\\
&&\mbox{} 
+ \left[\mbox{\small$10\over9$} D^2 
- \mbox{\small$4\over3$} DF - 2F^2 
- \mbox{\small$1\over3$} {\cal C}^2\right]\, V_{us}
- \left[2 (D/3+F)\gamma\right]\, V_{ss}
\\
M_{\Sigma^*}^{(b)} &=& 
(\mbox{\small$1\over9$} {\cal C}^2 - \mbox{\small$10\over81$} {\cal H}^2) 
[V_{ud} + V_{ds} + V_{su}] 
\nonumber\\
&&\mbox{}
-\mbox{\small$10\over27$} {\cal H}\gamma' [V_{uu} + V_{dd} + V_{ss}] \,,
\end{eqnarray}
where 
\begin{equation}
V_{us} = {M_{us}^3 \over 16 f^2 \pi} \,, {\rm etc.}
\end{equation}
Note that we have rewritten $\alpha$ and $\beta$ 
in terms of $D$ and $F$ using Eq.~(\ref{eq:Qparameters}).

These results for $M^{(b)}$ show that the $M_\pi^3$ terms 
are present in QQCD. Some are proportional to $D$ and $F$, i.e. of the
same general form as those in QCD, while others, proportional to $\gamma$
and $\gamma'$ are quenched artifacts. To give some idea of the effect of
quenching, we quote the QCD result for $\Lambda$ \cite{bkm}
\begin{equation}
M_\Lambda^{(b)}({\rm QCD}) = \left[ 2 D^2 + {\cal C}^2 \right] V_\pi
- \mbox{\small$2\over3$}\left[(D^2 + 9 F^2) + {\cal C}^2\right] V_K
- \mbox{\small$2\over3$} D^2 V_\eta \,.
\end{equation}
There is no obvious correlation between quenched and full results.

The diagrams of Fig.~(\ref{fig:feynman}c) are obtained from those
just considered by inserting an $m_0^2$ or $\alpha_\Phi$ vertex on
the meson line. We find 
\begin{eqnarray}
M_\Sigma^{(c)} &=&  
\mbox{\small$1\over2$}
\left[4 F^2\,W(u,u) - 4(D-F)F\,W(u,s) + (D-F)^2\,W(s,s)\right]
\nonumber\\
&&\mbox{} 
+ \mbox{\small$1\over9$}
{\cal C}^2 \left[W(u,u) - 2 W(u,s) + W(s,s)\right] \,,
\\
M_\Lambda^{(c)} &=& 
\mbox{\small$1\over2$}
\left[(4D/3\!-\!2F)^2\, W(u,u) - 2(4D/3\!-\!2F)(D/3+F)\, W(u,s)
\right.
\nonumber\\ 
&&\mbox{} \qquad \left.+ (D/3\!+\!F)^2\, W(s,s)\right]
\\
M_{\Sigma^*}^{(c)} &=& 
\mbox{\small$5\over162$} {\cal H}^2 
\left[W(u,u) + W(d,d) + W(s,s) + 2 W(u,d) + 2 W(d,s) + 2 W(s,u) \right]
\nonumber\\
&&\mbox{}
+\mbox{\small$1\over18$} {\cal C}^2 
\left[W(u,u) + W(d,d) + W(s,s) - W(u,d) - W(d,s) - W(s,u) \right] \,,
\end{eqnarray}
where 
\begin{equation}
W(a,b) = 
{2\over3} \left(
{ \alpha_\Phi (M_{aa}^5 - M_{bb}^5) - m_0^2 (M_{aa}^3-M_{bb}^3) 
\over 16 f^2 \pi (M_{aa}^2 - M_{bb}^2)} \right)
\,,
\end{equation}
and
\begin{equation}
W(a,a) = \lim_{a\to b} W(a,b) = 
{\mbox{\small$5\over3$}\alpha_\Phi  M_{aa}^3  
- m_0^2 M_{aa} \over 16 f^2 \pi} 
\,.
\end{equation}
These contributions are pure quenched artifacts. 
The $\alpha_\Phi$ term is of roughly the same form as
the ``conventional'' loop corrections proportional to $M_\pi^3$.
The $m_0^2$ term, however, is enhanced  by $m_0^2/M_\pi^2$ in the chiral
limit. It turns out, however, that the influence of
these terms on the curves of $M_{\rm bary}$ versus $M_{qq}^2$ is not
significant until quark masses substantially smaller than those used
in present simulations. We discuss this in detail in the companion paper.

It is simple to understand the origin of the enhanced loop contribution.
The second propagator in the loop makes the diagrams more infrared 
divergent than those in Fig.~(\ref{fig:feynman}b), and thus more
sensitive to the IR cut-off.
The contribution proportional to $m_0^2$ can be obtained, 
up to overall factors, by acting with
$m_0^2 (\partial/\partial M_{qq}^2)$ on the results from 
Fig.~(\ref{fig:feynman}b).
This replaces one power of $M_{qq}^2$ with $m_0^2$.
This is another example of the peculiar behavior of quenched quantities
in the chiral limit, previously seen in $M_\pi^2$
[see Eq.~(\ref{eq:1looppion})], $f_K$, $f_B$ and
various matrix elements \cite{bg,sharpe1,bg-lat92,sharpezhang}.

The diagrams of Figs.~(\ref{fig:feynman}d) and (e) yield 
further quenched artifacts
which are proportional to $m_q m_0^2 \log(M_{qq}^2)$. 
Without the hairpin vertex, these diagrams
give corrections proportional to $m_q M_\pi^2 \log(M_\pi^2)$,
which are suppressed in the chiral limit.
The insertion of the hairpin vertex makes the diagrams more
IR singular, replacing $M_\pi^2$ with $m_0^2$.
In quoting our results we have converted factors of $m_q$ into $M_{qq}^2$
using the lowest order formula. This is consistent since,
as we explain below, we are dropping terms of $O(m_0^4)$.

We find that
\begin{equation}
M_\Sigma^{(d)} = M_\Lambda^{(d)} = M_{\Sigma^*}^{(d)} = 0 \,.
\end{equation}
The cancelation between the two diagrams
is related to the simple flavor structure of graphs
involving the $m_0^2$ vertex---as illustrated in Fig.~(\ref{fig:sunset4}).
There is no such cancelation between the diagrams in (e), and we obtain
\begin{eqnarray}
M_\Sigma^{(e)} &=&  
- \mbox{\small$1\over9$}
{\cal C}^2 \left[ 
-4 b_F\,M_{uu}^2 + 2(b_D-b_F)\,M_{ss}^2 
- c'(2\,M_{uu}^2+M_{ss}^2) \right] \times
\nonumber\\
&&\mbox{}\qquad
[X(u,u) - 2 X(u,s) + X(s,s)] \,,
\\
M_\Lambda^{(e)} &=& 0 \,,
\\
M_{\Sigma^*}^{(e)} &=& 
\mbox{\small$1\over18$} {\cal C}^2 
\left[c' + 2(b_f-b_D/3)\right] (M_{uu}^2+M_{dd}^2+M_{ss}^2) \times
\nonumber\\
&&\mbox{}
\qquad \left[X(u,u) + X(d,d) + X(s,s) - X(u,d) - X(d,s) - X(s,u)\right]
\nonumber\\
&&\mbox{}
-\mbox{\small$1\over18$} {\cal C}^2 
 b_D (M_{uu}^2-M_{dd}^2) \left[X(u,u) - X(d,d) - 2\,X(u,s) + 2\,X(d,s)\right] 
\nonumber\\
&&\mbox{}
-\mbox{\small$1\over54$} {\cal C}^2 
b_D (M_{uu}^2+M_{dd}^2- 2 M_{ss}^2) \times
\nonumber\\
&&\mbox{} \quad 
\left[X(u,u) + X(d,d) - 2\,X(s,s) - 4\,X(u,d) +2\,X(u,s) +2\,X(d,s)\right] \,.
\end{eqnarray}
Here 
\begin{equation}
X(a,b) = {m_0^2 (M_{aa}^2\log(M_{aa}^2)-M_{bb}^2\log(M_{bb}^2)) 
	\over 16 f^2 \pi^2 (M_{aa}^2 - M_{bb}^2)} \,,
\end{equation}
and
\begin{equation}
X(a,a) = \lim_{a\to b} X(a,b) = 
{m_0^2 [\log(M_{aa}^2)+ 1] \over 16 f^2 \pi^2} \,.
\end{equation}
The last two terms in $M_{\Sigma^*}^{(e)}$ are symmetric under
permutations, despite appearances.

We can now explain 
why we consider only the diagrams of Fig.~(\ref{fig:feynman}). 
We are essentially carrying out a standard chiral expansion in $M_{qq}^2$,
supplemented by an expansion in $m_0^2/3$.
The next chiral corrections are $\sim M_{qq}^4 \log M_{qq}$, 
and thus suppressed by one power of $M_{qq}$ compared to the 
$M_{qq}^3$ terms that we keep. 
We also are dropping terms suppressed by additional powers of $m_0^2$. 
However, we keep contributions proportional to
$(m_0^2/3) M_{qq}^2 \log M_{qq}$, 
since they become important as we approach the chiral limit.
In particular, if $M_{qq}^2 < m_0^2/3$, the enhanced logarithm can
lead to $(m_0^2/3) \log M_{qq} \sim M_{qq}$, in which case
the logarithmic term should be kept (it is $\sim M_{qq}^3$), 
while it is consistent to drop higher powers of $(m_0^2/3) \log M_{qq}$.
In practice, in present simulations $M_{qq}^2 > m_0^2/3$,
so it is likely that $M_{qq}^4 \log(M_{qq})$ terms are significant.
Nevertheless, as simulations approach closer to the chiral limit,
our analysis will become more applicable.

A second reason for truncating the expansions is more practical.
The non-analytic corrections that we have calculated are those
that are determined unambiguously in terms of the 
lowest order coefficients in the chiral Lagrangian.
There is no dependence of unknown scales
$\sim \Lambda_\chi$, and no counterterms of the same order.
This is obvious for the terms proportional to $V$ and $W$, 
but is true also for those containing the $X(a,b)$---the 
scale of the logarithm cancels in the corrections $M_{\rm bary}^{(e)}$.
This is in contrast to meson masses and decay constants, 
e.g. Eq. (\ref{eq:1looppion}),
where the chiral corrections are proportional to 
$\log(M_{qq}^2/\Lambda_\chi^2)$ and thus contain an unknown scale.
This unknown scale does enter into $M_{\rm bary}^{(a)}$,
but can be absorbed into the meson masses as described above.

Finally, we recall that our calculation is done in the approximation
that $\Delta M \ll M_{qq}$, so that the octet and decuplet baryons
can be treated to first approximation as degenerate.
It would be straightforward in principle to extend our results to
larger $\Delta M$, although we have not carried out the calculation.
$\Delta M$ enters only in diagrams in which the internal baryon
is an octet while the external baryon is a decuplet, or vica versa.
These are the diagrams giving contributions proportional to ${\cal C}^2$.
The effects of $\Delta M\ne0$
can be included exactly by shifting the mass in the baryon
propagator in these diagrams, $k \cdot v \to k \cdot v \pm \Delta M$.
Some of the resulting integrals have been evaluated
in Ref.~\cite{jmlec}.
For example, for the second diagram of Fig.~(\ref{fig:feynman}b),
the factor of $M_\pi^3$ from the loop is replaced by an integral
whose expansion for small $\Delta M$ is
$M_\pi^3 - 3M_\pi^2 \Delta M \log(M_\pi)/\pi$.
In present simulations, the second term is a relatively small correction.

\section*{Acknowledgements}
We thank Maarten Golterman and Larry Yaffe for useful comments.

\end{document}